\documentclass{article}

\usepackage{arxiv}

\usepackage[utf8]{inputenc} 
\usepackage[T1]{fontenc}    
\usepackage{hyperref}       
\usepackage{url}            
\usepackage{booktabs}       
\usepackage{amsfonts}       
\usepackage{nicefrac}       
\usepackage{microtype}      
\usepackage{lipsum}
\usepackage{graphicx}

\usepackage{amsmath,amssymb}
\usepackage{enumerate}
\usepackage{longtable}
\usepackage{mathtools} 
\usepackage{multicol}
\usepackage{epstopdf}

\graphicspath{ {./images/} }

\title{Macroscopic Quantum Electrodynamics and  Density Functional Theory Approaches to Dispersion Interactions between Fullerenes}

\author{
 Saunak Das\thanks{Leibniz Institute of Photonic Technology (IPHT), Albert-Einstein-Str. 9, 07745 Jena, Germany.} \,\thanks{Current address: Department of Chemistry, The University of British Columbia, Vancouver, British Columbia, Canada. Stewart Blusson Quantum Matter Institute, The University of British Columbia, Vancouver, British Columbia, Canada.}\\
  Institute of Physical Chemistry (IPC)\\ Friedrich Schiller University Jena,\\ Helmholtzweg 4, 07743 Jena,\\ Germany. \\
   \And
 Johannes Fiedler \\
  Institute of Physics,\\ Albert-Ludwigs University of Freiburg,\\ Hermann-Herder-Str. 3, 79104 Freiburg, Germany.\\
  \texttt{johannes.fiedler@physik.uni-freiburg.de} \\
     \And
 Oliver Staufert \\
  Institute of Physics,\\ Albert-Ludwigs University of Freiburg,\\ Hermann-Herder-Str. 3, 79104 Freiburg, Germany.\\
     \And
 Michael Walter\thanks{FIT Freiburg Centre for Interactive Materials and Bioinspired Technologies, University of Freiburg, Georges-K\"ohler-Allee 105, 79110 Freiburg, Germany.}\,\,\thanks{Cluster of Excellence livMatS @ FIT -- Freiburg Center for Interactive Materials and Bioinspired Technologies, University of Freiburg, Georges-K\"ohler-Allee 105, 79110 Freiburg, Germany.}\,\,\thanks{Frauenhofer IWM, MikroTribologie Centrum $\mu$TC, W\"ohlerstrasse 11, 79108 Freiburg, Germany.} \\
  Institute of Physics,\\ Albert-Ludwigs University of Freiburg,\\ Hermann-Herder-Str. 3, 79104 Freiburg, Germany.\\
    \And
 Stefan Yoshi Buhmann \\
  Institute of Physics,\\ Albert-Ludwigs University of Freiburg,\\ Hermann-Herder-Str. 3, 79104 Freiburg, Germany.\\
  \And
 Martin Presselt\footnotemark[1]\,\,\thanks{Center for Energy and Environmental Chemistry Jena (CEEC Jena), Friedrich Schiller University Jena, Philosophenweg 7a, 07743 Jena, Germany.} \,\,\thanks{sciclus GmbH \& Co. KG, Moritz-von-Rohr-Str. 1a, 07745 Jena, Germany.}\\
  Institute of Physical Chemistry (IPC),\\Friedrich Schiller University Jena, \\Helmholtzweg 4, 07743 Jena, \\Germany. \\
  \texttt{martin.presselt@leibniz-ipht.de} \\
}

\begin{document}
\maketitle
\begin{abstract}
The processing and material properties of commercial organic semiconductors, for e.g. fullerenes is largely controlled by their precise arrangements, specially intermolecular symmetries, distances and orientations, more specifically, molecular polarisabilities. These supramolecular parameters heavily influence their electronic structure, thereby determining molecular photophysics and therefore dictating their usability as n-type semiconductors. In this article we evaluate van der Waals potentials of a fullerene dimer model system using two approaches: a) Density Functional Theory and, b) Macroscopic Quantum Electrodynamics, which is particularly suited for describing long-range van der Waals interactions. Essentially, we determine and explain the model symmetry, distance and rotational dependencies on binding energies and spectral changes. The resultant spectral tuning is compared using both methods showing correspondence within the constraints placed by the different model assumptions. We envision that the application of macroscopic methods and structure/property relationships laid forward in this article will find use in fundamental supramolecular electronics.
\end{abstract}


\section{Introduction}
 In addition to molecular structure,~\cite{Wuerther17,Habenicht16,Catalan,Hupfer19Langmuir} particularly the supramolecular structure essentially determines ensemble properties, such as the UV-vis absorption~\cite{Wuerthner2003,Das16,Hupfer19Chemistry} and emission~\cite{Habenicht16,Hupfer17} spectrum, charge transfers~\cite{Graham14}, conductivity~\cite{Rivnay2009,Noriega2013,OPV_Herrmann-PEDOT-PSS_APL2012} and further properties~\cite{Poelking2014,Das17}. Archetype examples are formations of H- or J-aggregates from dipolar dyes, which are accompanied by pronounced changes in the UV-vis absorption spectra.~\cite{Wuerther17} For a multitude of rather non-polar dyes~\cite{Gampe15,Ma08} aggregate formation and absorption spectra~\cite{Das16} depend on (supra)molecular polarisabilities. More specifically, these polarisabilities essentially influence the London dispersion interaction between the molecules and vice versa. Naturally, the molecular polarisabilities of dyes change upon photo-excitation and the accompanied redistribution of electron density.~\cite{Preiss17,Presselt09,Presselt10,SupraM-PhotoAnneal_Herrmann-JPCA2018} Therefore, both intermolecular binding and supramolecular structure of thin films must change upon photoexcitation. Depending on the ratio between dyes being photoexcited and being in the ground state at the same time the photo-induced supramolecular structural changes might reach a magnitude that significantly alters the (optoelectronic) material properties. Hence, it is of high interest to determine the dispersion interaction, or more precisely the van der Waals (vdW) potential, between molecules in the ground and in the electronically excited state.

 The estimation of dispersion interactions via purely microscopic methods is a huge issue for quantum chemical approaches, such as (time-dependent) density functional theory [(TD)DFT], because the resulting potentials couple to the polarisability, respectively dielectric function, which are quantities determined by macroscopic averaging or quantum mechanical expectation value methods.
 A complete description of these effects can be performed via macroscopic quantum electrodynamics (mQED). In this work, we will apply both methods to determine vdW-potentials and spectral properties of fullerene dimers, particularly in dependence on molecular rotation, translation and electronic excitation. We choose fullerenes as case study because their supramolecular structures are known to heavily influence photonic~\cite{Das16} and electronic~\cite{Das17,Shubina14} material properties.\cite{Nierengarten,OPV-SupraMol_Das-Review_JMatChemC2019,Theo-Mol_Sachse-OmegaTuning_JChemPhys2019}

Fullerenes (both C$_{60}$ and C$_{70}$ with their derivatives) find application in a plethora of different fields like nanomedicine~\cite{Dellinger13}, hydrogen storage~\cite{Pupysheva08}, bio-organics~\cite{Cheng2015} and photodynamic therapy~\cite{Mroz07}, but particularly as $n$-type semiconductors in several branches of organic electronics~\cite{Das17,Brabec08,Laquai15,Kaestner13}. The optical tunability of fullerenes can be ensured either by molecular variation or using supramolecular chemistry~\cite{Hollamby2014,Naire1600142}, as recently reviewed\cite{OPV-SupraMol_Das-Review_JMatChemC2019}. To date, non-bonded vdW-assemblies of fullerenes have received poor attention~\cite{Shubina14,Buhmann12,Nierengarten,OPV-SupraMol_Das-Review_JMatChemC2019} although their optoelectronic and material properties are known to depend on their precise molecular arrangements in the non-bonded assembly~\cite{Nakanishi10,OPV-SupraMol_Das-Review_JMatChemC2019,Das17}. This knowledge gap, particularly how vdW binding compares between electronic ground and excited states will be addressed in this work following a two-pronged approach: We will first discuss ground state vdW potentials of fullerene dimers as derived from DFT~\cite{PhysRev.140.A1133} and excited state vdW potentials as derived from TDDFT\cite{CASIDA} and finally analyse both properties  from the viewpoint of mQED. Thus, this manuscript recombines methods from DFT and QED.

Historically, both methods are based on fundamental quantum mechanics, but have been applied to different systems.
DFT enables to determine ground-state properties of a molecular or solid-state system. 
The extension to TDDFT also allows to address properties of excited states.
In contrast, quantum electrodynamics is the conjunction of quantum mechanics and classical electrodynamics that describes the interaction of light and matter from the viewpoint of the quantised electromagnetic field. 

The properties of quantum mechanical states are important for both theoretical approaches. The states involved are generally many body objects and the full range of their properties are described by their many-body wave functions. 
DFT gains its efficiency by the reduction of the complex wave function to the electron density which is sufficient to determine the ground state energy exactly. The Kohn--Sham approach further introduces effective single-particle molecular orbitals (MO) from which the exact electron density can be determined. While orbitals lack a clearly defined meaning, it is well known that they resemble measurable properties of many-body system if single particle properties such as photoelectron energies\cite{Kostko07,Stauffert17}, angular distributions\cite{Bartels13} or scanning tunneling microscopy images\cite{Walter07prl,Lin09} 
are considered.
The single particle picture can also fail for these observables, however\cite{Walter08njp,Dauth11}.
In contrast, the viewpoint of electromagnetic fields adopted in mQED is independent from the used description of molecular wave-functions, i.e. it can be applied to the many-particle as well as for the orbital picture case. In this work, we study the ground- and excited-state interactions of fullerenes by means of DFT and show that the predicted orientational dependence of the spectral response can be well fitted with analytical expressions derived with mQED. This offers the chance to considerably reduce the computational effort in studies of orientational effects and can be extended to macromolecules.

\begin{figure}
    \centering
        \includegraphics[width=0.7\columnwidth]{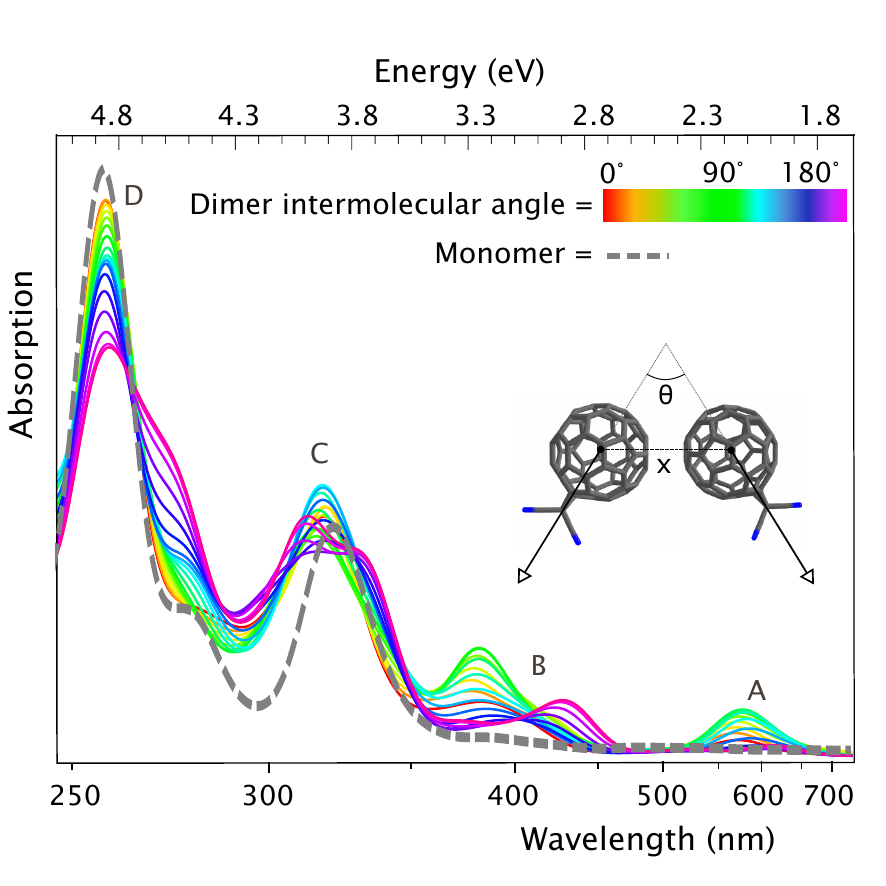}
    \caption{Simulated electronic absorption spectra for  $\rm{C}_{61}(\rm{CN})_2$ monomer and $\rm{C}_{61}(\rm{CN})_2$ C$_2v$ dimers as a function of intermolecular angle ($\theta$). Note that the center of masses of the individual fullerenes (separated by distance x) were unchanged upon rotation, thus keeping C$_2v$ symmetry and avoiding clashing of the fullerenes at any angle.}    \label{fig:5}
\end{figure}

In the manuscript, we start with a short introduction of the different methods which are applied to the system of two coupled C$_{60}$ derivatives, first briefly presenting the basics of the employed standard TDDFT approach and then outlining the essential aspects of the mQED approach in more detail. The subsequent results part is organised in sections on ground state and excited state interactions, comparing results from (TD)DFT and mQED for the dependence of the van der Waals interactions on the rotational angle between the two fullerenes. For completeness, a brief discussion of the (TD)DFT-derived dependence of the electronic fullerene states on the inter-fullerene distance is prepended to each section.

\section{Theories describing absorption peaks}

Within the Born-Oppenheimer approximation, absorption peaks in the visible can be identified with transitions between two electronic states $\left|\Phi_I\right\rangle$ and $\left|\Phi_J\right\rangle$. 
The corresponding transition frequency $\Omega_{IJ}$ equals to the energy difference between the two states~\cite{QM05}
\begin{equation}
    \hbar \Omega_{IJ} = \left\langle \Phi_I \right| \hat{\operatorname{H}}\left|\Phi_I\right\rangle -\left\langle \Phi_J \right| \hat{\operatorname{H}}\left|\Phi_J\right\rangle \,, \label{eq:QM_state_diff}
\end{equation}
where $\hat{\operatorname{H}}$ is the Hamiltonian of the molecular system and the states are eigenstates of it.

\subsection{Ab Initio Calculations}

Wave function theory would allow to calculate the electronic 
states 
$\left|\Phi_{I,J}\right\rangle$ and the corresponding
energies directly, but such a descriptions is practically impossible for the large number of electrons in C$_{60}$.
TDDFT is an efficient method to obtain the $\Omega_{IJ}$ as well as the corresponding oscillator strengths with 
manageable computational effort.\cite{Casida2012,Maitra2016} The price to pay is 
the introduction of an approximation for the unknown
exchange correlation potential. These approximations have been applied to an overwhelming number of systems ranging from single atoms over interacting molecules to solids and are thus well tested in the literature.\cite{Laurent2013} The approximations used
in our calculations are detailed in section \ref{sec:methods}.

The quantity of interest here, $\Omega_{IJ} ({\bf{r}},\phi,\theta)$, can thus be studied directly within
TDDFT, which leads to the transition energies and oscillator strengths of the coupled system from which Figure \ref{fig:5} is derived.
We will also use the orbital description, where the states entering Equation (\ref{eq:QM_state_diff}) are approximated by orbitals within the Kohn-Sham single particle picture, which is justified for the here-used generalised gradient approximation functional BP86.\cite{KohnSham1965,Theo-Mol_Sachse-OmegaTuning_JChemPhys2019} This viewpoint allows for a simplified analysis and assignment of transitions.

\subsection{Level Shift in Terms of mQED}\label{sec:mqed}

The orientational dependence of $\Omega_{IJ}$ can also be described via macroscopic quantum electrodynamics (mQED). This gives an alternative description of the interaction from a macroscopic viewpoint of the electromagnetic fields involved~\cite{Fiedler15,Brand15} and 
reveals the underlying reasons for the shifts.
The theory will be introduced and its connection to the shifts of spectral frequencies will be illustrated in what follows. 

We want to express the transition frequency of the joint molecular system in terms of the spectral properties of the monomers. This is enabled by the macroscopic quantum electrodynamics approach, which provides a perturbative solution of the expectation values in Eq.~(\ref{eq:QM_state_diff}) in terms of a single-molecule Hamiltonian $\hat{\operatorname{H}}_0$~\cite{QM05}
\begin{equation}
    \left\langle \Phi_{I} \right| \hat{\operatorname{H}}\left|\Phi_{I}\right\rangle \approx \left\langle \Phi_{I}\right| \hat{\operatorname{H}}_0\left|\Phi_{I}\right\rangle+ \left\langle \Phi_{I} \right| \hat{\operatorname{H}}_F\left|\Phi_{I}\right\rangle \,,
\end{equation}
with an effective field Hamiltonian $ \hat{\operatorname{H}}_F$ consisting of the electromagnetic fields created by the second molecule and similar for $J$. Thus, the results of the macroscopic quantum electrodynamical considerations are the energy level shifts of the corresponding states~\cite{0295-5075-110-5-51003}
\begin{equation}
    E_{I}= E^{(s)}_{I} + \Delta E_{I} \,,
\end{equation}
where $E_{I}$ denotes the total energy of the state in the interacting system, $E_{I}^{(s)}$ denotes the state's total energy for the single particle and $\hbar\Delta\Omega_{I}=\Delta E_{I}$ is the detuning induced by the environment consisting of the second particle. Electronic transitions contributed by overlapping $\pi$-orbitals are not treated using QED. However, intramolecular excitations can be described by macroscopic QED~\cite{Buhmann12a,Buhmann12b}, which treats both particles separately and calculates an environment-dependent Lamb shift~\cite{PhysRev.72.241,doi:10.1142/9383}. From this point of view, the interaction that has to be considered is the vdW energy of two anisotropic particles.~\cite{Eisenschitz1930}

\begin{figure}[t]
    \centering
    \includegraphics[width=0.6\columnwidth]{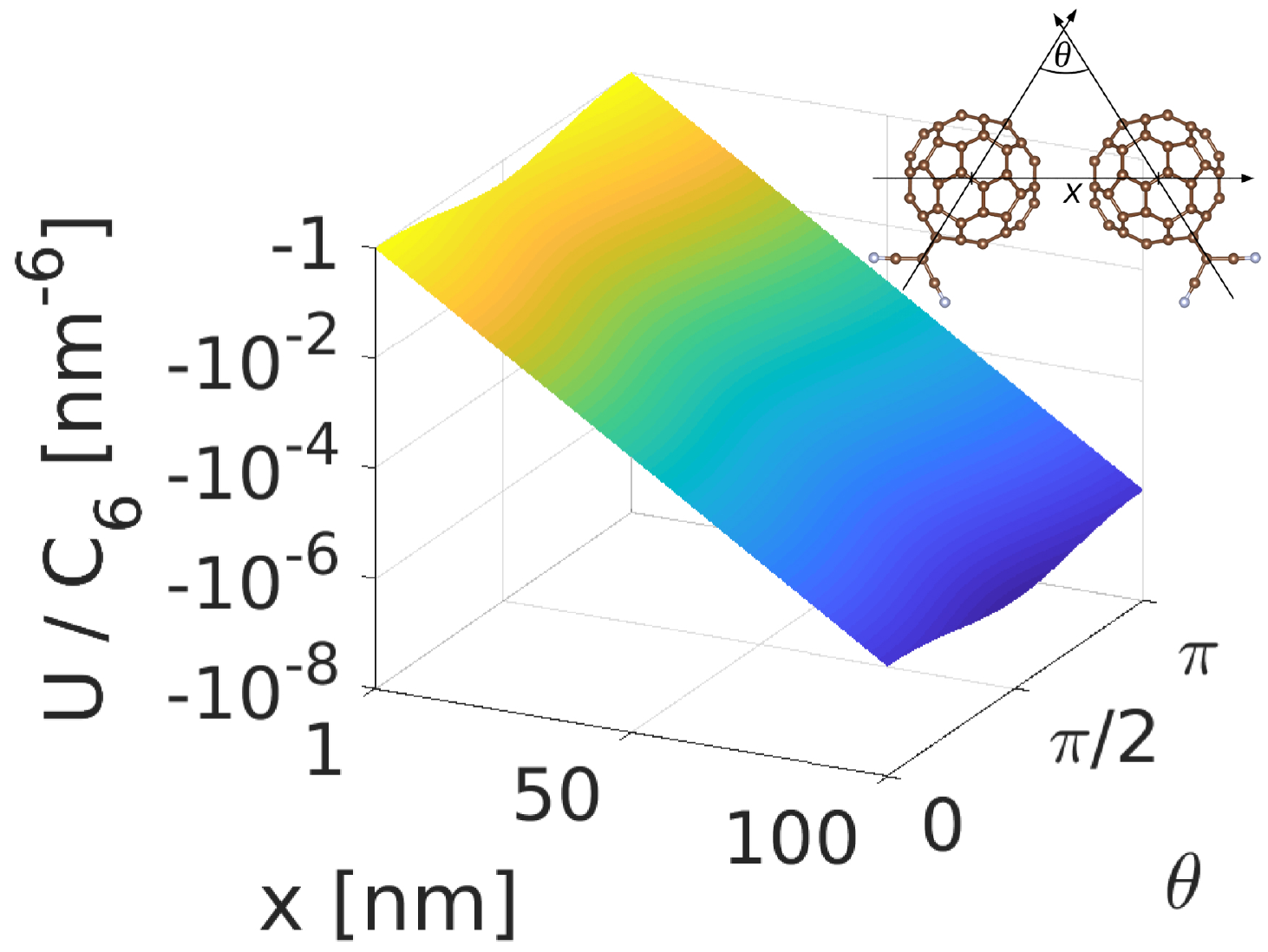}
    \caption{Sketch of the attractive vdW dispersion potential for two particles separated by a distance $x$ orientated by an angle $\theta$.}
    \label{fig:pot}
\end{figure}

The vdW potential between two neutral, but polarisable ground-state particles $A$ and $B$ located at ${\bf{r}}_A$ and ${\bf{r}}_B$ reads~\cite{Buhmann12a}
\begin{eqnarray}
U_{\rm vdW}({\bf{r}}_A,{\bf{r}}_B) = -\frac{\hbar \mu_0^2}{2\pi}\int\limits_0^\infty \mathrm d \xi \, \xi^4 \operatorname{tr}\left[ \alpha_A(i\xi) \cdot {\bf{G}}({\bf{r}}_A,{\bf{r}}_B,i\xi) \cdot \alpha_B(i\xi) \cdot {\bf{G}}({\bf{r}}_B,{\bf{r}}_A,i\xi)\right] \, ,\label{eq:Uvdw}
\end{eqnarray}
with the polarisabilities of the particles $\alpha_A(i\xi)$ and $\alpha_B(i\xi)$ evaluated at imaginary frequencies $i\xi$.
Equation~(\ref{eq:Uvdw}) is valid for ground as well as for excited states of the molecules characterised by the respective polarisabilities.
The considered arrangement is illustrated in the pictogram of Fig.~\ref{fig:pot}. Two anisotropic particles are orientated such that their principle axes form an angle $\theta$ and their centres  are separated by a distance $x$. The coordinate system is chosen in that way that the relative coordinate between the centres of both molecules are aligned long the $x$-axis, ${\bf{r}}_A-{\bf{r}}_B = (x,0,0)$. Eq.~(\ref{eq:Uvdw}) can be interpreted as a virtual photon with frequency $i\xi$ propagating from particle $A$ to $B$ where it interacts with its polarisability and is backscattered to particle $A$ and couples to its polarisability. The integral (which reduces to a sum for discrete modes) over these processes yields the vdW potential. The propagations are expressed by the Green's tensor ${\bf{G}}$ which is the solution of the vector Helmholtz equation for the electric field~\cite{Buhmann12a}.

The product of the Green's tensors with the polarisability of particle $B$, $ {\bf{G}}({\bf{r}}_A,{\bf{r}}_B,i\xi) \cdot \alpha_B(i\xi) \cdot {\bf{G}}({\bf{r}}_B,{\bf{r}}_A,i\xi)$, is closely related to the local mode density~\cite{PhotonsAtoms,doi:10.1142/9383} of the electric field at the position of particle $A$ in the presence of particle $B$. Hence, we consider the first particle in the field induced by the second one. In order to relate this method to the solution of the Schr\"odinger equation for the combined system comprising both objects, one approximates the field arising from the second particle with a mean field. This picture directly yields the limitation of this method, namely the spatial separation between both object necessary to distinguish between both objects. At the same time this is the reason for the restriction to the transitions around 4.8 eV; for the other transitions the overlap between the corresponding states is too large. This field in the presence of particle $B$ couples to the ground-state polarisability of  particle $A$, which yields the energy shift of the particle's ground state. Thus, we can write the ground-state detuning as
\begin{equation}
    \Delta E_0 = U_{\rm vdW} \, .
\end{equation}

Let us consider a specific transition between the ground state $J=0$ and an excited state $I=1$. Thus, the transition frequency reads as $\hbar\Omega_{10}=E_1-E_0$. Hence the energy shift that is observed in the presence of a second particle depends on the detuning of the involved orbitals
\begin{equation}
    \Delta\Omega_{10} =\frac{U_{\rm vdW}^1 -U_{\rm vdW}^0}{\hbar} \,, \label{eq:spec}
\end{equation}
where $U_{\rm vdW}^i$ denotes the vdW potential acting on the $i$-th electronic state, which can be obtained by using the respective polarisability in Eq.~(\ref{eq:Uvdw}). For excited particles the resonant transitions contribute as well as the non-resonant part to the vdW potential, which can be obtained via the Casimir--Polder potential~\cite{Buhmann12b}
\begin{equation}
    U_{\rm CP}({\bf{r}}_A) = -\frac{\mu_0}{3} \sum_{k<n} \omega_{nk}^2 \left|{\bf{d}}_{nk}\right|^2\operatorname{tr}\left[{\bf{G}}({\bf{r}}_A,{\bf{r}}_A,\omega_{nk})\right] \, , \label{eq:res}
\end{equation}
where the Green's function is given by the scattering at particle $B$ and can be obtained via the Hamaker approach \cite{Burger18}
\begin{equation}
    {\bf{G}}({\bf{r}}_A,{\bf{r}}_A,\omega) = \frac{\omega^2}{c^2\varepsilon_0}{\bf{G}}({\bf{r}}_A,{\bf{r}}_B,\omega)\cdot \boldsymbol{\alpha}_B(\omega) \cdot {\bf{G}}({\bf{r}}_B,{\bf{r}}_A,\omega) \, .
\end{equation}
We are particularly interested in the orientational dependency of energy shift determined by the corresponding vdW potential. To this end, we introduce the polarisability of an anisotropic particle as 
\begin{equation}
    \boldsymbol{\alpha}(i\xi) = \alpha_{zz}(i\xi)
    \operatorname{diag}(e,e,1) \, , \label{eq:pol}
\end{equation}
with a scalar effective eccentricity $e$ of the particle's polarisability~\cite{Fiedler18}
\begin{equation}
    e= \frac{ \int\alpha_{xx}(i\xi) \mathrm d \xi}{\int\alpha_{zz}(i\xi) \mathrm d \xi} \,,\label{eq:exxen}
\end{equation}
where the side group of the monomer is pointing into the $z$-direction.
The eccentricity of the electronic ground state can be evaluated directly from TD-DFT calculations.
In case of excited states we obtain the eccentricity from fits of the dependence of the transition energy on the intermolecular angle as detailed below. The orientational dependence follows by replacing the polarisabilities in Eq.~(\ref{eq:Uvdw}) with the ones from the rotated framework
\begin{equation}
    \alpha_{A,B}(i\xi) \mapsto {\bf{R}}(\theta) \cdot \alpha_{A,B}(i\xi)\cdot {\bf{R}}^{-1} (\theta)\, ,
\end{equation}
with the rotation matrix
\begin{eqnarray}
{\bf{R}}(\theta) =\begin{pmatrix}
\cos\theta & 0 & \sin\theta\\
0 & 1 & 0 \\
-\sin\theta & 0 & \cos\theta
\end{pmatrix} \,,
\end{eqnarray}
and the rotation angle is defined as depicted in Fig. \ref{fig:5}. By using the non-retarded vacuum Green's tensor~\cite{Buhmann12a}
\begin{equation}
    {\bf{G}}(x,x',i\xi) =-\frac{c^2}{4\pi\xi^2 \left(x-x'\right)^3} \begin{pmatrix}
    -2 & 0 & 0 \\
    0 & 1 & 0 \\
    0 & 0 & 1
    \end{pmatrix}\,,
\end{equation}
and choosing one particle to be located at the origin ($x'=0$), the ground-state vdW potential reads~\cite{Fiedler18}
\begin{eqnarray}
\label{eq:UvdW}
U_{\rm vdW}  \left(x,\theta \right) =U_{\rm vdW}^{\rm iso}(x)\left[1+ \frac{1}{4} \left(\cos^{4}\theta +2\cos^2\theta + 2 \right) \left({e}^{2} -1 \right)+ \frac{1}{2} \left(\cos^2  \theta - \cos^4\theta \right) \left(e-1 \right) \right] \,, \label{eq:rotvdw}
\end{eqnarray}
with the well-known van der Waals potential for isotropic particles
\begin{equation}
    U_{\rm vdW}^{\rm iso}(x) =-\frac{C_6}{x^6}\,,\label{eq:Uvdw_iso}
\end{equation}
with $C_6 = \frac{3\hbar}{16\pi^3\varepsilon_0^2} \int\limits_0^\infty \operatorname{tr}\alpha_A(i\xi) \alpha_B(i\xi)\mathrm d \xi$, which is in our case $C_6 = 64048\, \rm{eV\text{\AA}}^6$, which is consistent with the vdW interaction between two buckyballs~\cite{PhysRevA.85.042513,Andersson_1999,doi:10.1063/1.2035589,PhysRevLett.109.233203,PhysRevLett.79.3873} and also in agreement with DFT simulations. The resulting potential landscape is plotted in Fig.~\ref{fig:pot}. Due to the high $C_6$-coefficient for the van der Waals interaction, this interaction dominates the intermolecular forces which is orders of magnitude larger than the Debye and Keesom forces. This result explains the spatial and the orientational dependence by the spatial variable $x$ and the angle $\theta$, respectively. The vdW potential for the excited state remains the same with a changed $C_6$-coefficient that separates additively into two parts: one for the resonant~(\ref{eq:res}) and one for the non-resonant~(\ref{eq:Uvdw}) contribution. In this case, it is sufficient to derive the interaction with the ground state particle, because the unknown excited state enters the further analysis as a fitting parameter. Finally, this methods allows the description of the absorption peaks for the dimer based on the monomer's parameters (polarisability $\alpha$ and eccentricity $e$) and the geometric arrangement of both molecules. For a prediction also the parameters for the excited states (polarisabilities $\alpha_n$ and eccentricities $e_n$) are required.

\subsection{Effective particle separation}
\begin{figure}
    \centering
    \includegraphics[width=0.7\columnwidth]{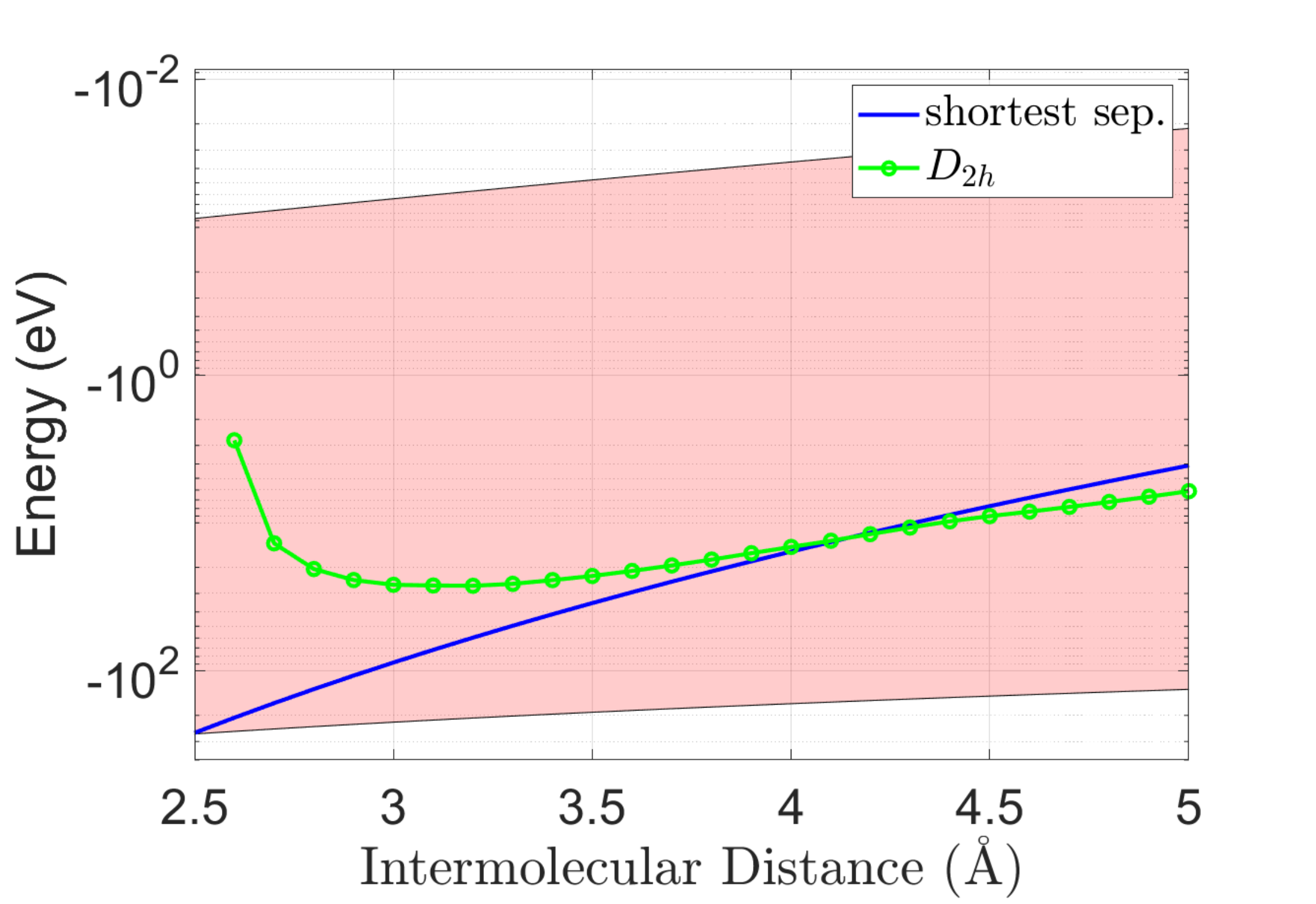}
     \caption{Comparison of the intermolecular energies obtained via DFT [$D_{2h}$ (marked green line)] and via mQED [Eq.~(\ref{eq:Uvdw_iso}) (upper boundary of the red area), Eq.~(\ref{eq:Uvdw_iso}) corrected to the shortest separation (blue line) and the Hamaker approach~(\ref{eq:Ham}) (lower boundary of the red area)].}
    \label{fig:distances}
\end{figure}

The derived van der Waals potential~(\ref{eq:Uvdw_iso}) deviates from the DFT simulations due to the point-particle assumption. The distance between both particles ($\approx 10.5${\AA}) is comparable with their extension (diameter $2r\approx 7${\AA}). In principle, there are two ways to improve the agreement between both theories: (i) an expansion to higher-order multipoles~\cite{PhysRevB.2.3371}; (ii) an effective treatment via dielectric spheres~\cite{PhysRevLett.99.170403}. Both ways are equivalent but require lengthy calculations. A lower limit of the range of potentials can be provided by considering both particles as dielectric spheres which interact via the Casimir potential. By applying Clausius--Mossotti relation~\cite{doi:10.1021/acs.jpca.7b10159,Jackson} to the obtained polarisability
\begin{equation}
    \alpha(\omega) = 4\pi\varepsilon_0 r^3 \frac{\varepsilon(\omega)-1}{\varepsilon(\omega)+2}\,,
\end{equation}
a dielectric function representing the spheres can be obtained. This result can be used further to calculate the Hamaker constant of this system
\begin{equation}
    H = \frac{3 k_{\rm B} T}{2} \sum_{m=0}^\infty {}' \mathrm{Li}_3\left[\left(\frac{\varepsilon(i\xi_m)-1}{\varepsilon(i\xi_m)+1}\right)^2\right] \,,
\end{equation}
with the polylogarithmic function $\rm{Li}_3$ and the Matsubara frequencies $\xi_m=m\,2\pi k_{\rm B} T/\hbar$. This results in a Hamaker constant for the system of $H=3.6619 \cdot 10^{-20}\,\rm{J}\approx 2286\,\rm{eV}$. Thus, the energy between both spheres can be approximated as~\cite{israelachvili1985intermolecular}
\begin{equation}
E_{\rm Ham}(x)= - \frac{H}{6x} \frac{r}{2}\,, \label{eq:Ham}
\end{equation}
depending on the intermolecular distance $x$ and corrected due to the curvature of the spheres $r$. The results are shown as the lower boundary of the red area in Fig.~\ref{fig:distances}. Its upper bound is given by the van der Waals potential~(\ref{eq:Uvdw_iso}). 

This transition goes hand in hand with a reduction of the distance between both particles from the centre-of-mass distance to the shortest separation between both spheres. Hence, an effective size-corrected van der Waals potential can be constructed by shifting the separation accordingly by twice the radius of the sphere. The result is depicted by the blue line in Fig.~\ref{fig:distances} compared with the DFT result given by the green line. It can be observed that this shifted potential approximates the long distance tail of the DFT simulation very well and further considerations of higher-order multipoles are redundant. For this reason, we apply this reduction of the particle separation to our further analysis. 

\section{Fullerene's Ground State vdW Potential from DFT and mQED}
Most fullerenes targeted for commercial applications are asymmetrically substituted to facilitate dissolution in various solvents and chemical processing~\cite{Troshin09,Wang12}. Furthermore, the absorption spectra of fullerenes heavily depend on the high symmetry of the buckyball, which gets disturbed upon the introduction of substituents~\cite{Wang12,Orlandi02,Hare1991394}. To mimic asymmetric substitution but allow for theoretical treatment within point group theory, we will focus on the recently introduced $(\rm{CN})_2\rm{C}_{61}$ model~\cite{Das16}. Within DFT, we considered the dispersion interaction between the fullerenes by applying Grimme's D3-correction~\cite{Grimme06,Goerigk11}.

 \begin{figure}[t]
    \centering
    \includegraphics[width=0.7\columnwidth]{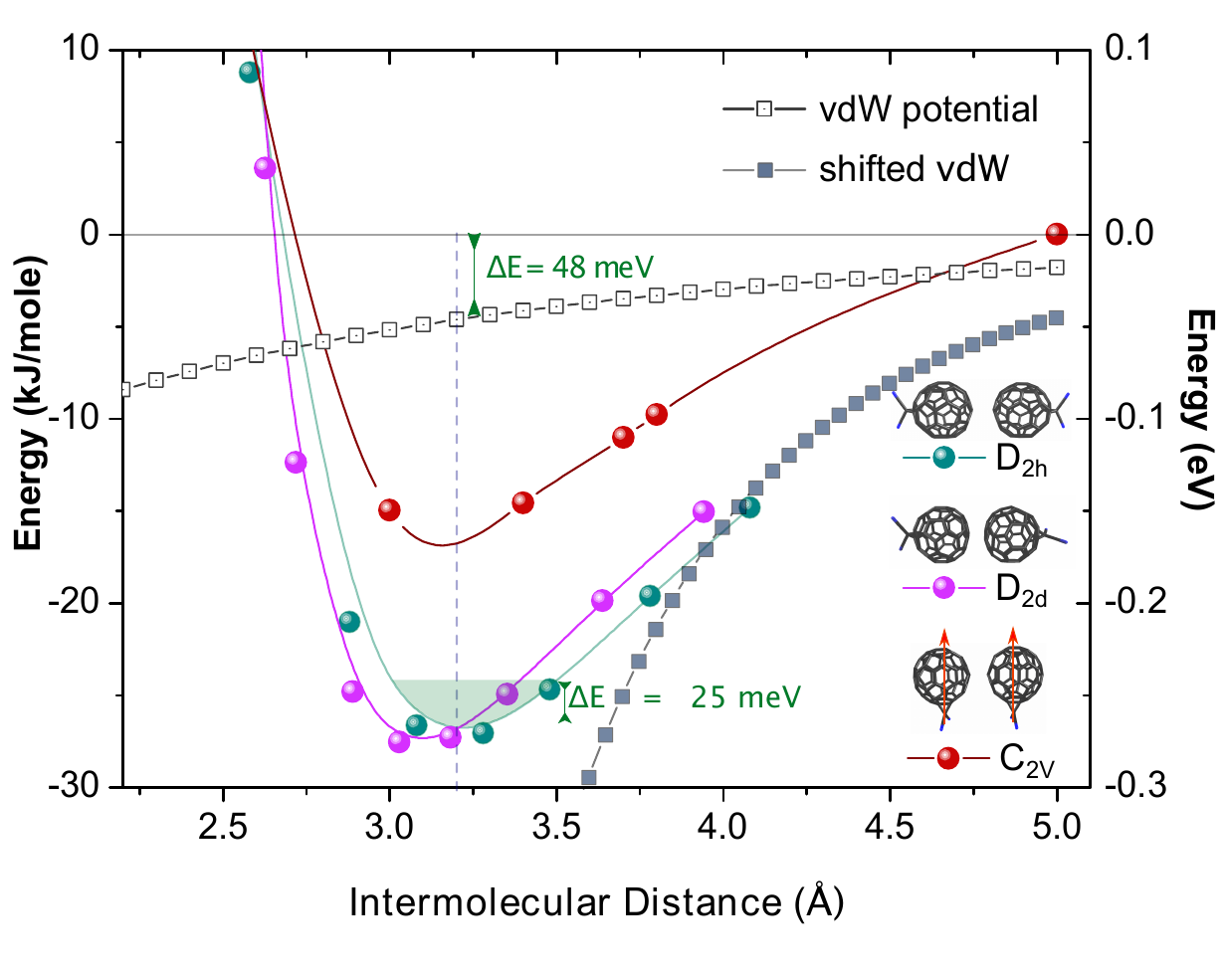}
    \caption{Binding profiles for $(\rm{CN})_2\rm{C}_{61}$ dimers of different symmetry derived via DFT calculations and the van der Waals (vdW) potential according mQED, Eq.~(\ref{eq:rotvdw}) with $\theta=180^\circ$.
    %
    Zero on the $y$-axis refers to non-interacting monomers. For both D$_{2h}$ and C$_{2v}$-dimers the (9,10) $\pi$-bonds of the closest naphthalene like moieties on each fullerene are oriented parallely whereas in the D$_{2d}$-dimer these surface $\pi$-bonds are oriented perpendicular to one another. The shaded 25 meV represents the thermal energy at room temperature.}
    \label{fig:1}
\end{figure}

\subsection{Dependence on Intermolecular Distance}
Ground state van der Waals potentials derived via DFT are shown in Fig.~\ref{fig:1}. The dissociation tails of these vdW-potentials are determined by keeping the distance fixed and by relaxing all other degrees of freedom. These simulations are additionally restricted by keeping the initially determined point group for fixed distances, namely point groups D$_{2d}$, D$_{2h}$ and C$_{2v}$, see Fig.~\ref{fig:1}.  Shallow binding energy minima are observed between 2.9-3.2 {\AA} with energies of 27.4, 26.9 and 16.8~kJ/mol for dimers with point groups D$_{2d}$, D$_{2h}$ and C$_{2v}$, respectively. Interestingly, the dispersion interaction provides almost the whole binding energies. Considering the fullerene derivative to be a molecular dipole, the intermolecular parallel dipole-dipole interaction is expected to be more repulsive in C$_{2v}$ than D$_{2h}$ or D$_{2d}$ dimers, see Fig.~\ref{fig:1}. Therefore, both D$_{2h}$ and D$_{2d}$ dimers are stabilised by 10.6 and 10.1~kJ/mol larger binding energies, respectively, than the C$_{2v}$ dimer. The dipole-dipole interactions operating in all investigated dimers is attributed as reason for binding energies being lower than the energy of 50~kJ/mol reported by Shubina et al.\cite{Shubina14}.

In Figure~\ref{fig:1}, the DFT-derived binding profiles are compared to the mQED-derived van der Waals potential between the eccentric C$_{60}$ derivatives (according Eq.~(\ref{eq:rotvdw}) using $\theta=180^\circ$ and $C_6 = 64048\, \rm{eV\text{\AA}}^6$, as derived above). The fact that the mQED-vdW potential is shallower than the dissociation/association part of the DFT-derived potential might either be attributed to too low $C_6$ coefficients if the binding potential shall be described only within the vdW model (see ref.~\cite{PhysRevLett.109.233203}) or to an transition from vdW effect between point-like objects to the Casimir effect~\cite{Dalvit} between extended spheres as the separation of both molecules is similar to their extension at the equilibrium distance. This transition to the Casimir effect can be approximated in the first order by introducing a reduced separation of the particles as explained above. By comparing the potentials for separations larger than 4{\AA} obtained via DFT and mQED, we find that this reduction is approximately described by the shortest separation between both particles. In Figure~\ref{fig:1}, it can be observed further that the potential shifted by the diameter of a fullerene molecule follows the tail of the long range attraction. Thus, we can conclude that the main difference between the van der Waals potential and the energies obtained by DFT is caused by the missing of short range repulsion.

\begin{figure}[t]
    \centering
    \includegraphics[width=0.7\columnwidth]{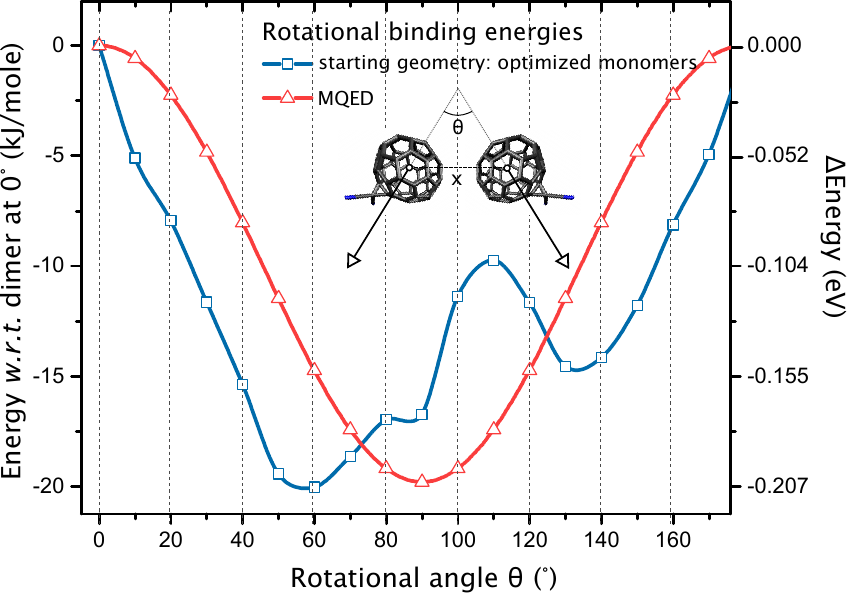}
    \caption{Impact of the rotation onto the total energy of $\rm{C}_{61}{\rm(CN)}_2$-dimer. Plotted is the relative energy difference with respect to parallel fullerenes as a function of intermolecular angle as determined by single point calculations on dimers created from optimised monomers at varied angles. Note that the centre of masses of the individual fullerenes were unchanged upon rotation, thus keeping C$_{2v}$ symmetry and avoiding clashing of the fullerenes at any angle. The $y$-axis is defined by setting the energy of the 0 degree dimer at 0 kJ/mole. The mQED-derived energy profile is derived using Eq.~(\ref{eq:ff}).}
    \label{fig:2}
\end{figure}

\subsection{Dependence on Intermolecular Angle}

\begin{figure}
    \centering
    \includegraphics[width=0.7\columnwidth]{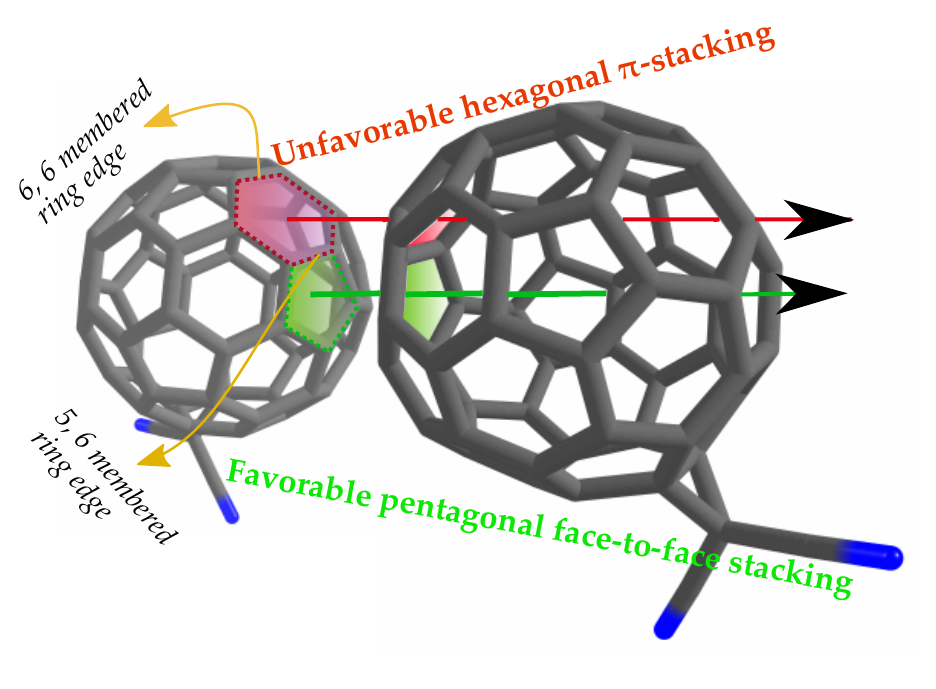}
    \caption{Sketch depicting interacting rings between the fullerene interstitial spaces. Pentagonal face-to-face interaction is favoured, while hexagonal $\pi$-stacking is unfavourable as discussed in text.}
    \label{fig:3}
\end{figure}

The DFT-derived potential for varied intermolecular angles at constant distances of 3.2 {\AA} are shown in Fig.~\ref{fig:2}. The intermolecular angle $\theta$ is varied in 10$^{\circ}$ steps from 0 to 180$^{\circ}$.
Interestingly, two minima are noticed: one at ~60$^{\circ}$ which is about 5 kJ/mole deeper in energy than the second minimum at ~135$^{\circ}$. Similarly, dimers at $\theta=0^{\circ}$ and $\theta=180^{\circ}$ are both located at energetic maxima, while a third crest is obtained at 110$^{\circ}$. These features of the rotational energy profile are assigned to fullerene ring/edge interactions as follows: 
\begin{itemize}
    \item[a)] At 60$^{\circ}$, the aliphatic 5-membered rings interact face to face and are oriented parallely. Because these 5-membered rings formally lack $\pi$-electrons, only favourable vdW interactions persist in this case (also see Fig.~\ref{fig:3}). 
\item[b)]	At 110$^{\circ}$ the edges between 5- and 6-membered rings are closest to each other, thus giving rise to destabilisation by 10~kJ/mol w.r.t. to the  favourable dimer at 60$^{\circ}$. 
\item[c)]	At 135$^{\circ}$ the aromatic six-membered $\pi$-rings interact face to face, hence causing destabilisation by 6~kJ/mol w.r.t. to the energetically most stable dimer at 60$^{\circ}$.
\item[d)]	Finally, at 180$^{\circ}$ the edge between two six-membered rings interact, thus yielding destabilisation by 19~kJ/mol w.r.t. energetically stable dimer at 60$^{\circ}$.
\end{itemize}

The model describing the change of the intermolecular interaction via mQED according to Eq.~(\ref{eq:ff}) reproduces the main features in the total energy curve estimated via DFT fairly well. The deviations between both models are due to the approximations which have been used, such as the restriction to the dipole-dipole interaction in the mQED description leading to the potential minimum at 90$^{\circ}$ instead of the DFT-derived global minimum at 60$^{\circ}$ or local minimum at 130$^{\circ}$, as shown in Fig.~\ref{fig:2}. The two DFT-derived minima at 60$^{\circ}$ and 130$^{\circ}$ are due to $\pi$-stacking between 5-membered and 6-membered rings, respectively, (Fig.~\ref{fig:3}).

The above discussed local potential thus provides physical insights into the relative stability order of ring versus edge interactions which can therefore be extracted using DFT only: 5-5 ring interaction > 6-6 ring interaction (+6~kJ/mol) > 5-6 edge interaction (10~kJ/mol) > 6-6 edge interaction (+19~kJ/mol).

The deviations in the rotational potential energy profiles resulting from mQED and DFT are due to neglection of intermolecular electronic repulsions in mQED and can be reduced by considerations of quadrupoles and octopoles.

\begin{figure*}[t]
    \centering
    \includegraphics[width=0.9\textwidth]{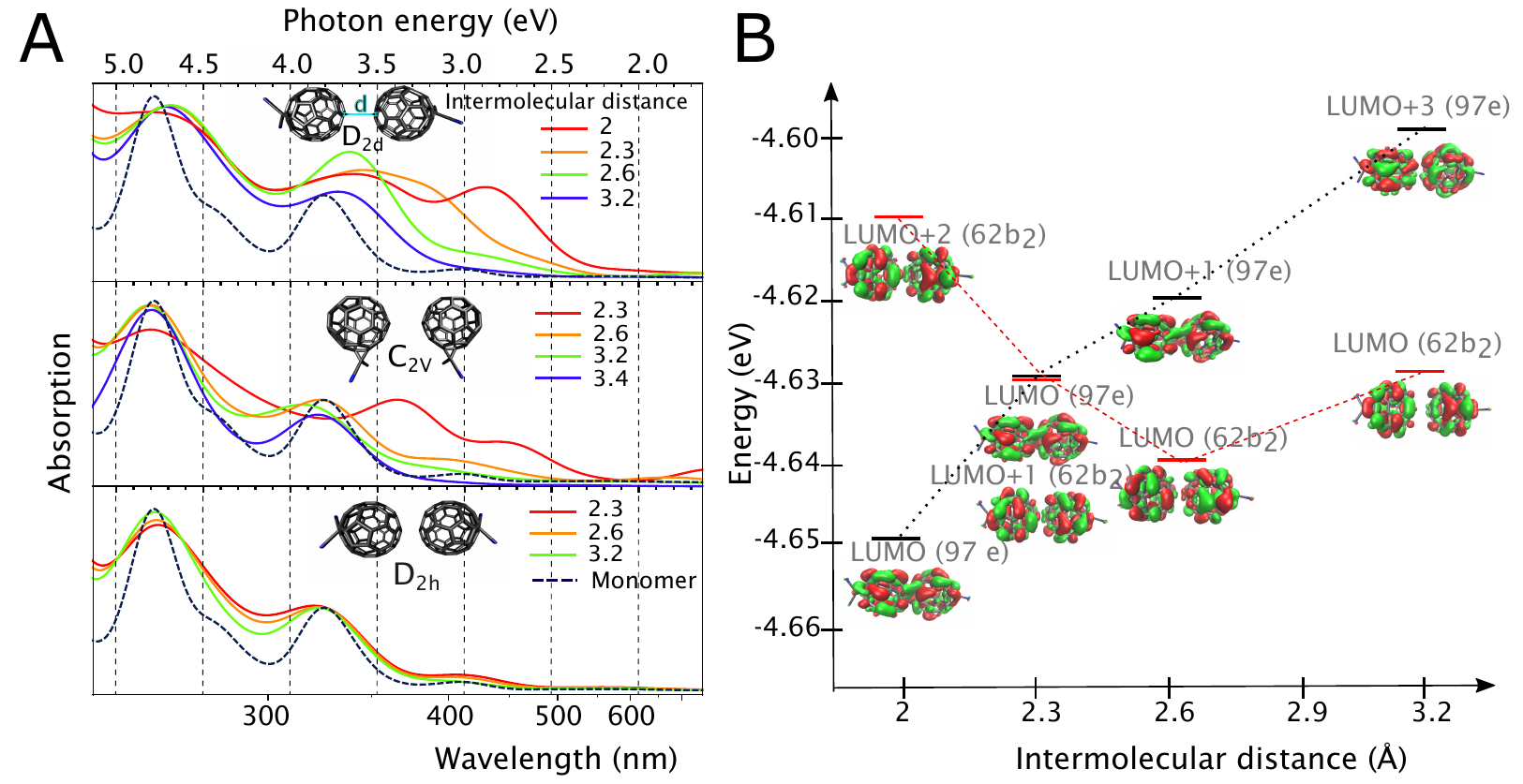}
    \caption{(A) TD-DFT-derived UV-vis absorption spectra (broadened via sums of Gaussian functions) depending on the intermolecular distance. (B) Exemplary change in energetic relation and order between lowest unoccupied molecular orbital  (LUMO) energies upon intermolecular distance variation. The two different LUMOs belong to the irreproducible representations $e$ (labelled $97_e$) and $b_2$ (labelled $62_{b2}$), hence facilitating their tracking (via symmetry and shape) in the reordered LUMO energy lists at each individual dimer geometry along varying the intermolecular distance.}
    \label{fig:4}
\end{figure*}

\section{Influence of Fullerene-Fullerene Interactions on Electronic Excitations}
\subsection{Dependence on Intermolecular Distance}

Reviewing the major absorption characteristics of the $\rm{C}_{61}(\rm{CN})_2$-monomer spectrum from a molecular viewpoint (see Fig.~\ref{fig:4} and ref.~\cite{Das16} for comparison between theoretical and experimental  spectra referring to systematically varied molecular assemblies), is an essential prerequisite for the following discussion on TDDFT-calculated fullerene dimer spectra. In short, the fullerene monomer spectrum is characterised by two intense UV-absorptions at 256 nm and 326 nm and weak vis-absorption. The 256 nm-peak involves two major transitions, similar to the two prominent transitions to nT$_{1u}$-excited states ($n=6,7$) in $I_h$-symmetric C$_{60}$~\cite{Orlandi02,Hare1991394}. The peak at 326 nm is a superposition of medium-intense and weak transitions, which is similar to the transition to the 3T$_{1u}$ state in $I_h$-C$_{60}$ (335 nm). A sharp peak at 425 nm is characteristic for fullerene derivatives with $[6,6]$-bridged carbons~\cite{Hare1991394}. The weak vis-absorption originates from several transitions that are symmetry-forbidden in $I_h$-C$_{60}$, but gain intensity because of Herzberg--Teller intensity stealing~\cite{Orlandi02}. Furthermore, combination bands contribute to these transitions~\cite{Orlandi76}, which comprise totally symmetric modes and non-totally symmetric Jahn--Teller active modes that partially favour these otherwise forbidden transitions~\cite{Orlandi02}.

VdW-dimerisation, reduction of symmetry, and $\pi$-orbital delocalisation increases absorption in the otherwise weak visible part of the spectra while the UV-region gets broadened due to origin of new transitions~\cite{Montero12}. The forbidden part of the spectra contains intermolecular delocalised states, which show a distance dependence~\cite{Montero12, Das16}. Therefore, as shown in Fig.~\ref{fig:4}, the TDDFT-calculated vis-absorption of D$_{2d}$ and C$_{2v}$ dimers gain intensity progressively as the distance is systematically reduced to 2 \AA. In case of the centrosymmetric D$_{2h}$ dimer, these symmetry-forbidden transitions do not escalate upon distance reduction, which otherwise would be a violation of Laporte's selection rule~\cite{Laporte:25}. An intensity gain of the same absorption band was predicted by Montero-Alejo et al. using the CNDOL method for a non-inversion-symmetric assembly of pristine C$_{60}$ molecules, indicating good agreement between CNDOL and TDDFT methods.\cite{Montero12}

To explain the reason of progressively higher vis-absorption upon distance reduction we take the case of dimer D$_{2d}$. Upon intermolecular distance shrinking at constant point group symmetry, i.e. all rotational degrees of freedom are frozen, interactions between the molecular $\pi$-orbitals lead to the formation of new bonding and anti-bonding orbitals that are delocalised over the whole dimer. Naturally, electronic excitations mediated by delocalised unoccupied orbitals will have lower transition energies than ones mediated by the non-delocalised orbitals (e.g. same dimer with spatially separated monomers). Thus, transitions associated with these stabilised MOs are bathochromically/hypsochromically shifted relative to those transitions involving energetically degenerate MOs. This leads to spectral broadening upon increasing intermolecular interaction.

In Fig.~\ref{fig:4}~(B), the example of the lowest unoccupied molecular orbital (LUMO) (62b${_2}$) (-4.63 eV) and the LUMO+3 (97$e$) orbitals (-4.6~eV) for the intermolecular distance 3.2~\AA \, is presented. LUMO+3 (97$e$) is $\pi$-$\pi$ delocalised and gets reduced in energy upon intermolecular distance reduction by 20~meV at 2.6~\AA. On further distance reduction, 97$e$ is systematically stabilised to become the LUMO for the fullerenes separated by 2~\AA, which is in total 50~meV more stabilised than that for 3.2~\AA. Seemingly, electronic transitions to LUMO shifts bathochromically as intermolecular distance diminishes. On the contrary, LUMO (62b$_2$) (-4.3~eV) at intermolecular distance 3.2~\AA \, shifts first away and then towards vacuum, thus contributing to an overall destabilisation of 20~meV at 2~\AA. Such delocalised states are frequent in the visible part of D$_{2d}$'s absorption spectrum explaining why  low energy absorption rise upon distance reduction. We note that orbital overlap between the interacting fullerenes does not appear when the interstitial hexagonal rings interact face-to-face. Such interactions are unfavourable due to electrostatically repulsive vdW surfaces for face-to-face $\pi$-stacking, which is why e.g. benzene dimers are A-B stacked.

\subsection{Dependence on Intermolecular Angle}

In Figure~\ref{fig:5}, the TDDFT-derived absorption spectra for $\rm{C}_{61}(\rm{CN})_2$ dimers at systematically increasing intermolecular angles are presented and compared with the monomer $\rm{C}_{61}(\rm{CN})_2$ absorption spectrum. Within the energy window considered (1.6~eV to 5~eV), all dimer spectra appear broadened and are strongly vis-absorbing w.r.t. the monomer UV-vis spectrum. The $\rm{C}_{61}(\rm{CN})_2$ monomer displays strong UV absorption with $\lambda_{max}$ at 256 nm (4.8~eV) followed by a second maxima at 326 nm (3.8~eV) and weak vis-absorption signals, thus reproducing the typical absorption features of the C$_{60}$ absorption spectrum \cite{Orlandi02}.

In case of the dimers, two distinct TDDFT-derived spectral changes are observed with increase in intermolecular angle (depicted by a colour bar in Fig.~\ref{fig:5}) between the fullerenes: a) peak splitting (peaks C and D) and, b) reduced oscillator strengths (A, B, C and D). Furthermore, a spectral shift of peak B was noticed. The prime electronic transitions in the presented absorption spectra are given by four different regions of interest (first between 1.8 and 2.3 eV, second between 2.7 and 3.3 eV, third between 3.6 and 4.2 eV and the fourth between 4.5 and 5 eV) as depicted in Fig.~\ref{fig:5}. Of these, the first two low energy regions (peaks A and B) are primarily populated by transitions involving delocalised states with $a_1$ and $b_2$ irreducible representations. The first singlet transition is of $a_1$-type occurring from HOMO to LUMO at 2.03 eV.

\begin{figure*}
    \centering
    \includegraphics[width=0.8\textwidth]{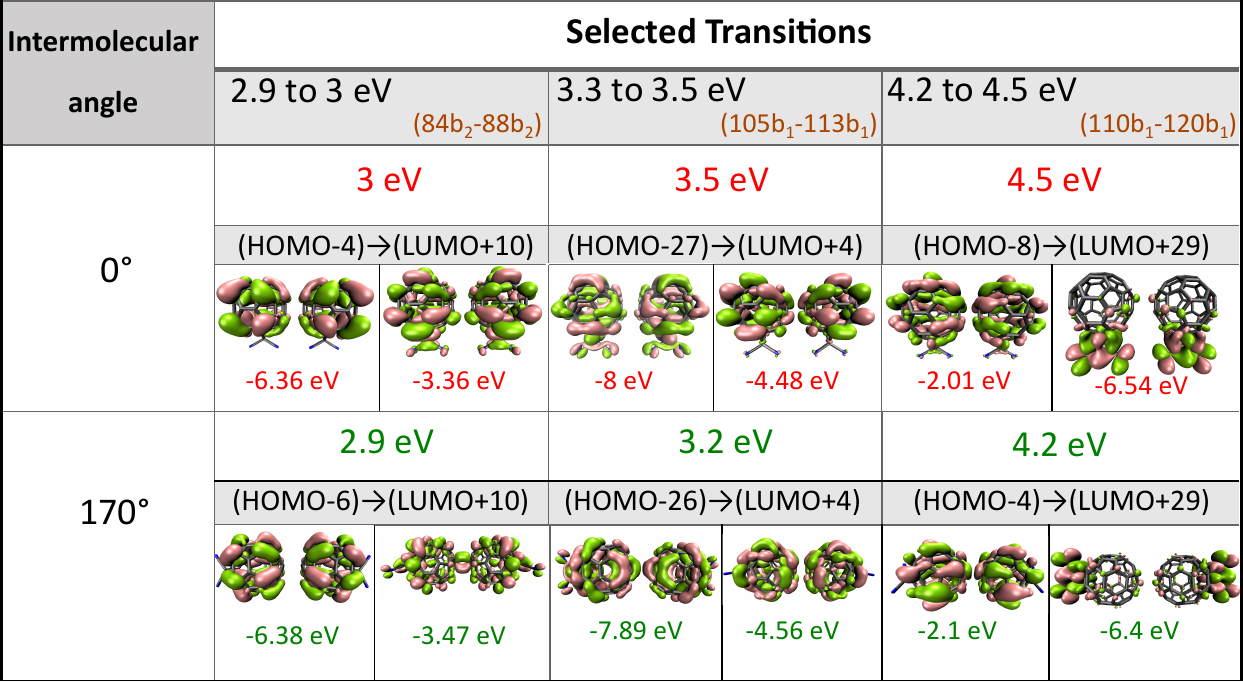}
    \caption{Illustration of orbitals involved in the selected transitions in the spectral regions of interest for the minimum ($\theta=0^\circ$) and large intermolecular angle ($\theta=170^\circ$) C$_{\textit{2v}}$ geometries. Note that contributing orbitals for the largest intermolecular angle ($\theta=180^\circ$) geometries are not plotted here as it belongs to D$_{\textit{2h}}$ point group resulting in incomparable orbital shapes \textit{w.r.t} plotted C$_{\textit{2v}}$ orbitals.}
    \label{fig:table}
\end{figure*}

To investigate and assign specific TDDFT-derived spectral features first the most intense electronic transitions for the four above-mentioned spectral regions were identified. It was observed that the excitation energy for these intense transitions are virtually constant upon rotation and are contributed by several molecular orbitals having small mixing coefficients (see section 1 of Supporting Information (SI)). Electronic transitions which are close in energy w.r.t. these intense transitions however show intermolecular angle dependency. These transitions can be well understood by one occupied (HOMO-x) and unoccupied (LUMO+y) orbital because of their high mixing coefficients (see Figure 1,2 and 3 of SI). Figure~\ref{fig:table} shows these electronic transitions, the involved orbitals with their corresponding energies for the initial (0$^{\circ}$) and final (170$^{\circ}$) geometries. It is necessary to bear in mind that the 180$^{\circ}$ geometry belongs to a D$_{2h}$ point group and is not considered in the present discussion but is duly discussed in earlier reports.\cite{Das16} From Fig.~\ref{fig:table}, it is clear that the energies of the tabulated transitions decrease as the intermolecular angles increase. Looking at the dominating orbitals for each transitions it seems reasonable that this overall change in excitation energy is a result of energetic stabilisation of the constituting orbitals. Here, orbital stability stems from delocalisation (\textit{cf.} LUMO+10 for transitions between 2.9 to 3~eV), or  destructive interactions as orbitals with opposite signs come closer (\textit{cf.} occupied orbitals for transitions between 3.3 to 5~eV and between 4.2 to 4.5~eV).  Overall, the TDDFT derived absorption spectra indicate that buckyball rotation in fullerene dimers can cause maximum spectral shifts of about ~300~meV.

In order to compare TDDFT to mQED results one needs to obtain the dynamical polarisability tensor of the single particle as input parameters entering the macroscopic quantum electrodynamics approach via the scalar polarisability~(\ref{eq:pol}) and the corresponding eccentricity~(\ref{eq:exxen}). Evaluating these parameters for the considered C$_{60}$ derivatives one finds an energy at the particles distance of $\Delta E_0=C_6 x^{-6}= 54 \,\rm{eV}$. 
The eccentricity of the electronic ground state of the monomer obtained from TDDFT is $e_0=0.83$. Figure~\ref{fig:pot} illustrates the orientational dependence of the vdW potential according to Eq.~(\ref{eq:UvdW}). It can be observed that the potential oscillates, which is caused by the rotational symmetry of the rotation angle.

The result from the macroscopic description is a correction function of the vdW potential with respect to the orientation of the particle~\cite{Fiedler18} in Eq.~(\ref{eq:UvdW}),
\begin{eqnarray}
f \left(e,\theta \right) =1+ \frac{1}{4} \left(\cos^{4}\theta +2\cos^2\theta + 2 \right) \left({e}^{2} -1 \right) + \frac{1}{2} \left(\cos^2  \theta - \cos^4\theta \right) \left(e-1 \right)  \,, \label{eq:ff}
\end{eqnarray}
independent from the particle's distance $x$. Due to this functional dependence of the vdW interaction on the rotational effects and the particle's eccentricity, one finds a relative impact of the rotations in the order of ten percent compared to the anisotropy effect of the particle. In Fig.~\ref{fig:5} the expected absorption spectrum is shown, where one observes four characteristic  peaks, labelled with A, B, C and D from low to higher energies. By varying the angle between both monomers one can see that the lowest peak (A) strongly increases in intensity, whereas the next characteristic peak (B) around 3 eV changes its position and amplitude with respect to the angle variation. The higher frequency peaks (C around 3.8 eV and D around 4.8 eV) show a low variation in the position but a stronger variance in the amplitude. One observes further the evolution of side bands for the higher frequency peaks. For the comparison, we will focus on the frequency shifts for the higher energy peaks B, C and D, where we focus on the most contributing orbitals.

\begin{figure}[t]
    \centering
    \includegraphics[width=0.7\columnwidth]{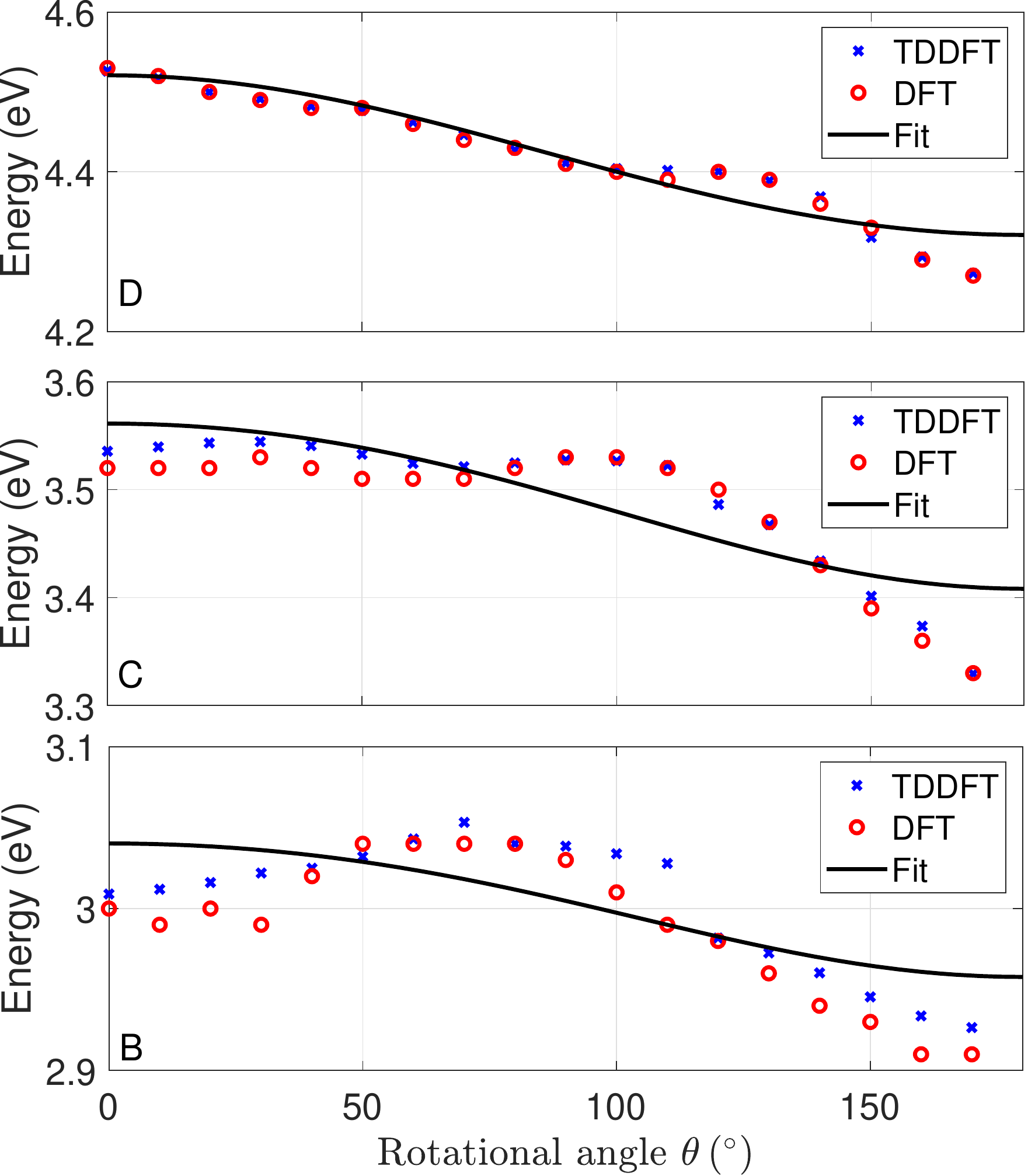}
    \caption{Fit of the considered transitions (labelled with B, C and D corresponding to Fig.~\ref{fig:5}) with blue markers denoting the TD-DFT data used for fitting. The red circles represent the corresponding orbital energy differences obtained via DFT according to Fig.~5 in SI.
    }
    \label{fig:fit}
\end{figure}

In order to compare the shifts of the transition energies estimated via TD-DFT with the macroscopic quantum electrodynamics approach, we map the transition onto an effective two-level system, see Fig.~\ref{fig:fit}.  
Inserting the relation~(\ref{eq:ff}) into the spectral shift~(\ref{eq:spec}), we can write the spectral energy shift for the transition between two states $\Phi_0$ and $\Phi_1$ as
\begin{equation}
    \Delta\Omega_{10}=\frac{\Delta E_1 f(e_1,\theta)-\Delta E_0 f(e_0,\theta)}{\hbar}\,, \label{eq:fitstate}
\end{equation}
where $\Delta E_{0,1}$ describes the energy shift of ground and excited state, respectively, and $e_{0,1}$ denotes the corresponding eccentricities. The ground-state is described by the vdW potential~(\ref{eq:UvdW}). Previous considerations of the vdW force for an excited particle showed that its functional dependence is the same compared to the ground state case for a two-level system~\cite{Buhmann12b}, but with changed parameters that are in this case the $C_6^e$-coefficient and the eccentricity defining the fitting parameters. By considering the mixing of several states, one has to take into account that each level contributes with its own strength and eccentricity. As mentioned above, the referencing ground-state system is characterised by the vdW energy at DFT-derived equilibrium distance $\Delta E_0=54 \,\rm{eV}$ (see Fig. 1) and the ground-state polarisability has an eccentricity of $e=0.83$. To this end, the fitting parameters according to the model~(\ref{eq:fitstate}) are the eccentricity of the excited state $e_1$ and the energy at equilibrium distance $\Delta E_1$.

\begin{table}[b]
    \centering
    \begin{tabular}{c|ccc}
        transition &  eccentricity 
         & $\Delta E_1 \,[\rm{eV}] $ & fit. agr.\\\hline
D & 0.82 & 50.9 & 92\% \\
C & 0.84  & 56.9 & 74\% \\
B & 0.84  & 56.1 & 60\% 
    \end{tabular}
    \caption{Results of the fitting routine for the three tracked transitions estimated via TDDFT simulations ($\left|\Phi_n\right\rangle \rightarrow \left|\Phi_{n-1}\right\rangle$), the corresponding eccentricities of the excited state, and the fitting agreement (fit. agr.), which is the coefficient of determination $R^2$.}
    \label{tab:states}
\end{table}

The results of the fitting routine are plotted in Fig.~\ref{fig:fit} and the relevant parameters are given in Table~\ref{tab:states}. It can be observed that the quantum electrodynamical model matches the rotational variation of the transition energy very well for higher transition energies. The comparison of the model prediction with the calculated energies shows a general representation of the results, see Fig.~\ref{fig:fit}. 
One can conclude from the fit that the repulsive contribution of the Lennard-Jones potential almost cancels for high transition energy states as evident from the very good fitting agreement. This implies that the Lennard-Jones component is isotropic for that particular state. For the states with lower transition energies the fitting agreement is somewhat reduced. This is attributed in parts to anisotropic Lenard-Jones repulsion and in parts to the neglect of higher multipole moments in our mQED description.

\section{Conclusion and Outlook}

In this work we have studied on calculation of intermolecular interactions in the electronic ground and excited states, which essentially determine supramolecular structures, hence being key for material properties. While the determination of distance-dependent energy profiles is rather straightforward, e.g. via fitting  Lennard-Jones potentials to quantum chemically derived energies, accounting for rotations to explore the full energy hypersurfaces might be computationally demanding using grid-based methods.\cite{Theo-SupraM_Sachse-EnergyScan-JCC2018,Theo-SupraMol_Metz-AutoGenInterMolPot_JCTC2016} Here we combine (time dependent) density functional theory and macroscopic quantum electrodynamics via the orbital coupling described by van der Waals interactions to provide energies for distinct grid-points of the energy hypersurface and analytic functions to describe the latter, respectively.

As a model system we have used dimers of asymmetrically substituted fullerenes. We showed that the mQED predictions, involving only the DFT-derived monomer properties polarisability and eccentricity, are well suited to fit the (TD)DFT-derived dimer intermolecular translational and rotational potentials. A reduction of the number of fitting parameters would require further knowledge about the polarisabilities for excited molecular states. For the special case of fullerene-dimers, the changes in TD-DFT transition energies upon rotation was evinced to originate from Kohn--Sham orbital energy differences obtained from DFT calculations. The results obtained via TD(DFT) and mQED via van der Waals potentials match well even though indications of the transition from van der Waals to Casimir regime has been observed. An analysis of this effect will be object of future investigations.

In a subsequent work, we will apply the here-presented approach to molecules featuring a significantly higher structural anisotropy, i.e. also a larger eccentricity, thus yielding a stronger dependence of the total energy on intermolecular rotational angles. For such a case we will also explore the limits of coarsening the spatial and rotational grids underlying the mQED fits, to estimate potentials of reducing computational effort for (TD)DFT calculations.

\section{Method Details}
\label{sec:methods}
Quantum chemical structure optimisations, and calculations of absorption spectra on fullerene monomers were performed using density functional theory and its time-dependent derivative implemented in Turbomole~\cite{Ahlrichs1989165,Furche14}. Here the GGA (generalised gradient approximation) BP86 exchange-correlation functional, TZVP~\cite{Weigend05} with MARI-J approximation~\cite{Furche14} was applied. Combinations of such kind gives chemically acceptable geometries, electron density distributions and spectroscopic parameters at reasonable computational cost~\cite{Presselt09,Beenken13,Presselt08,Presselt10b,Beenken14,Preiss15,Theo-Por_Presselt-TPPBasicity_PCCP2015}. BP-86 functional is also usually used in computing C$_{60}$ absorption spectrum in n-hexane~\cite{MENENDEZPROUPIN201472}. For geometry optimisation of the dimers we also employed the D3-dispersion-correction~\cite{C2CP24096C}. This yields an acceptable description of the dimers~\cite{Shubina14}, even though the vdW coefficients between fullerenes are abnormally large~\cite{PhysRevLett.109.233203}. 
Molecular orbitals were rendered by single point calculations on these individual geometries. 2700 electronic states were computed for each of these structures, disposing an equal number of 900 states among the A$_1$, B$_1$, B$_2$ irreducible representations. 

The DFT  implementation from the
GPAW-package~\cite{gpaw1,gpaw2} was used for the calculation of the eccentricity of the monomer in the electronic ground state. The electron density and the Kohn--Sham orbitals were represented on real space grids. The simulation grid had a spacing of 0.25 \AA{} for the orbitals and was  ensured to cover at least 4 \AA{} of space around each atom in the molecule. Zero boundary conditions were applied  outside of the simulation box. The exchange correlation functional was approximated by the GGA  devised by Perdew, Burke and Ernzerhof (PBE)~\cite{PBE}. TD-DFT was applied in Casida's linear response formalism~\cite{CASIDA,Walter08jcp}. Hereby from the 300 orbitals included in the simulation, those within an transition energy range of 20 eV are considered in the linear response calculation. From this we obtain excitation energies  and transition dipole moments that define the polarisability tensor $\alpha$ \cite{CASIDA} used in Eq.~(\ref{eq:exxen}).

\section*{Acknowledgements}
We gratefully acknowledge support from  the German Research Council (grant BU 1803/6-1, S.Y.B. and J.F., BU 1803/3-1, S.Y.B.), the Research Innovation Fund by the University of Freiburg (S.Y.B., J.F., M.W.),  the Freiburg Institute for Advanced Studies (S.Y.B.), and the Bundesministerium f\"ur Bildung und Forschung (BMBF FKZ 03EK3507, M.P.).

\bibliographystyle{unsrt}  


\newpage
\section{Supplemental Information}
\subsection{Method of Transition Tracking}
The TDDFT derived absorption spectra of C$_{61}$(CN)$_2$-dimers comprise of thousands of transitions populated by several molecular orbitals. To explain changes in the excitation energies, we identified how change in intermolecular angles affect the nature and shape of these orbitals. Therefore, major transitions with similar contributing orbitals at different spectral regions were identified as discussed below:
\begin{itemize}
    \item[a)]The absorption spectra of the dimer with 0 degree angle is most identical, and the one with 170 degree rotation angle is most dissimilar to the monomer spectra. 
\item[b)] Thus, the 4 most intense transitions of the 0 degree dimer are first selected and the most dominant orbitals describing these transitions are noted. It is found that several orbitals contribute to these transitions. An example, for the transition with highest oscillator strength is provided below:
78 orbitals (39 occupied and 39 unoccupied) are needed to describe this intense transition. New contributing orbitals appear for other rotation angles, but the excitation energy is virtually constant. Thus we do not track these transitions. Instead, we track those transitions which can be completely described by changes in orbital energies as described below.

\begin{table}[h!]
    \centering
    \begin{tabular}{ll}
Excitation energy / eV: &    4.560859854720713\\
 Excitation energy / nm: &                  271.8440117780391\\
 Oscillator strength: & \\
    mixed representation: &               0.1431629574598253    \end{tabular}
    \caption{Excitation energy for the fullerene-dimers.}
\end{table}

Dominant contributions:
\begin{longtable}{ccccc}
occ. orbital  & energy / eV &  virt. orbital    & energy / eV  & $|{\rm{coeff.}}|^2\times100$\\
 110 b1   &          -6.54 &         120 b1 &            -2.01&       18.1 \\
103 b1    &         -9.05  &       113 b1    &         -4.48   &    15.2 \\
101 a1    &         -9.26  &       112 a1    &         -4.70&       14.0 \\
110 b1    &         -6.54  &       119 b1    &         -2.21 &       6.9 \\
110 a1    &         -6.64  &       120 a1    &         -2.02  &      6.4 \\
109 a1     &        -7.47  &       116 a1    &         -2.63   &     4.6 \\
108 b1  &           -7.42  &       116 b1    &         -2.46   &     3.1 \\
83 a2    &         -6.37 &          90 a2    &         -2.37&        1.9 \\
106 a1    &         -7.71 &        115 a1    &         -3.26 &       1.7 \\
110 b1     &        -6.54 &        117 b1    &         -2.42  &      1.4 \\
84 b2       &      -6.36  &        90 b2     &        -2.41    &    1.3 \\
110 a1       &      -6.64 &        119 a1    &         -2.41    &    1.3 \\
108 a1  &           -7.53 &        115 a1    &         -3.26     &   1.2 \\
110 a1   &          -6.64 &        118 a1    &         -2.42      &  1.1 \\
105 b1    &         -8.00 &        114 b1    &         -3.43&        1.1 \\
110 a1     &        -6.64 &        117 a1    &         -2.55 &       1.0 \\
102 b1  &           -9.08 &        113 b1    &         -4.48  &      1.0 \\
106 b1   &          -7.53 &        115 b1    &         -3.20   &     0.9 \\
110 a1    &         -6.64 &        116 a1    &         -2.63    &    0.6 \\
107 a1     &        -7.66 &        116 a1    &         -2.63     &   0.6 \\
83 b2   &          -6.40  &        90 b2     &        -2.41       & 0.5 \\
109 a1    &         -7.47 &        118 a1    &         -2.42       & 0.5 \\
85 a2      &       -6.22  &        91 a2     &        -2.28&        0.5 \\
109 b1  &           -7.30 &        117 b1    &         -2.42&        0.5 \\
111 b1   &          -6.14 &        118 b1    &         -2.25 &       0.4 \\
85 b2     &        -6.23  &        91 b2     &        -2.31   &     0.4 \\
80 a2      &       -7.45  &        89 a2     &        -2.43    &    0.4 \\
78 a2       &      -8.40  &        86 a2     &        -4.32     &   0.4 \\
111 a1       &      -6.43 &        119 a1    &         -2.41     &   0.4 \\
80 b2   &          -7.54  &        89 b2     &        -2.50       & 0.4 \\
85 b2    &         -6.23  &        92 b2     &        -1.48        &0.4 \\
107 a1    &         -7.66 &        113 a1    &         -4.56&        0.4 \\
85 a2      &       -6.22  &        92 a2     &        -1.47  &      0.4 \\
109 b1  &           -7.30 &        115 b1    &         -3.20  &      0.4 \\
106 a1      &       -7.71 &        112 a1    &         -4.70   &     0.4 \\
108 a1  &           -7.53 &        117 a1    &         -2.55 &        0.3 \\
78 b2     &        -8.40  &        86 b2     &        -4.39   &     0.3\\ 
    \caption{Orbital energies for the fullerene-dimer.}

\end{longtable}
Below we select few transitions which change upon rotation, given by changes in absorption spectra.  
\item[c)] The tracking procedure considers those transitions which can be described by shift in orbital energies contributing to shifts in absorption spectra, prominent in naked human-eye. The transitions tracked fall in the following zones:
\begin{enumerate}
    \item Isosbectic point at 4.5 eV,
\item Peak-splitting at 3.6 eV, and
\item Shift in peak-maxima from 3 eV to 2.6 upon increased rotation angle, giving the impression of a second isosbectic point at, 2.8 eV.
\end{enumerate}

\item[d)] We select 3 intense transitions, at the vicinity of the energy regions in above-mentioned: a, b or c and describe them in the light of both TD-DFT and MQED. 

\begin{itemize}
    \item[(I)]  Isosbestic Point at 4.5 eV

    \begin{longtable}{ccccccc}
    \parbox{2cm}{Angle (${}^\circ$)} &
\parbox{2cm}{Excitation
Energy (eV) @TDDFT} &
\parbox{2cm}{Unoccupied
Orbital (UO) [eV] @DFT} &
\parbox{2cm}{Occupied Orbital (OO) [eV] @DFT} &
\parbox{2cm}{$\Delta$Orbital Energy
(UO-OO) [eV]}&
\parbox{2cm}{Mixing
Coefficient (\%)} &
\parbox{2cm}{Oscillator Strength}\\\hline
0 & 4.52612 & -2.01 & -6.54 & 4.53 & 65.2 & 5.82E-02\\
10 & 4.51728 & -2.03 & -6.55 & 4.52 & 53.2 & 1.92E-01\\
20 & 4.49976 & -2.05 & -6.55 & 4.5 & 39.4 & 2.80E-02\\
30 & 4.49134 & -2.06 & -6.55 & 4.49 & 78.8 & 2.58E-02\\
40 & 4.48221 & -2.07 & -6.55 & 4.48 & 76.8 & 1.15E-02\\
50 & 4.47688 & -2.07 & -6.55 & 4.48 & 64.9 & 1.84E-03\\
60 & 4.46216 & -2.08 & -6.54 & 4.46 & 81.4 & 2.93E-02\\
70 & 4.44641 & -2.09 & -6.53 & 4.44 & 82.3 & 1.40E-02\\
80 & 4.42726 & -2.09 & -6.52 & 4.43 & 92.9 & 0.9E-02\\
90 & 4.41096 & -2.1 & -6.51 & 4.41 & 90.1 & 6.53E-03\\
100 & 4.40398 & -2.1 & -6.5 & 4.4 & 83 & 0.5E-02\\
110 & 4.40184 & -2.1 & -6.49 & 4.39 & 84 & 0.43E-02\\
120 & 4.40045 & -2.09 & -6.49 & 4.4 & 83.7 & 3.34E-03\\
130 & 4.38893 & -2.09 & -6.48 & 4.39 & 57.4 & 1.38E-03\\
140 & 4.36878 & -2.1 & -6.46 & 4.36 & 73.1 & 2.45E-03\\
150 & 4.31782 & -2.1 & -6.43 & 4.33 & 44.1 & 1.84E-03\\
160 & 4.29355 & -2.1 & -6.39 & 4.29 & 72.5 & 8.87E-05\\
170 & 4.27282 & -2.1 & -6.37 & 4.27 & 40.3 & 4.46E-04\\\hline
    \caption{Angle dependence of the excitation energies and occupied and unoccupied orbitals for the isosbestic point at 4.5 eV.}

    \end{longtable}


        \begin{figure}
        \centering
        \includegraphics[width=0.7\textwidth]{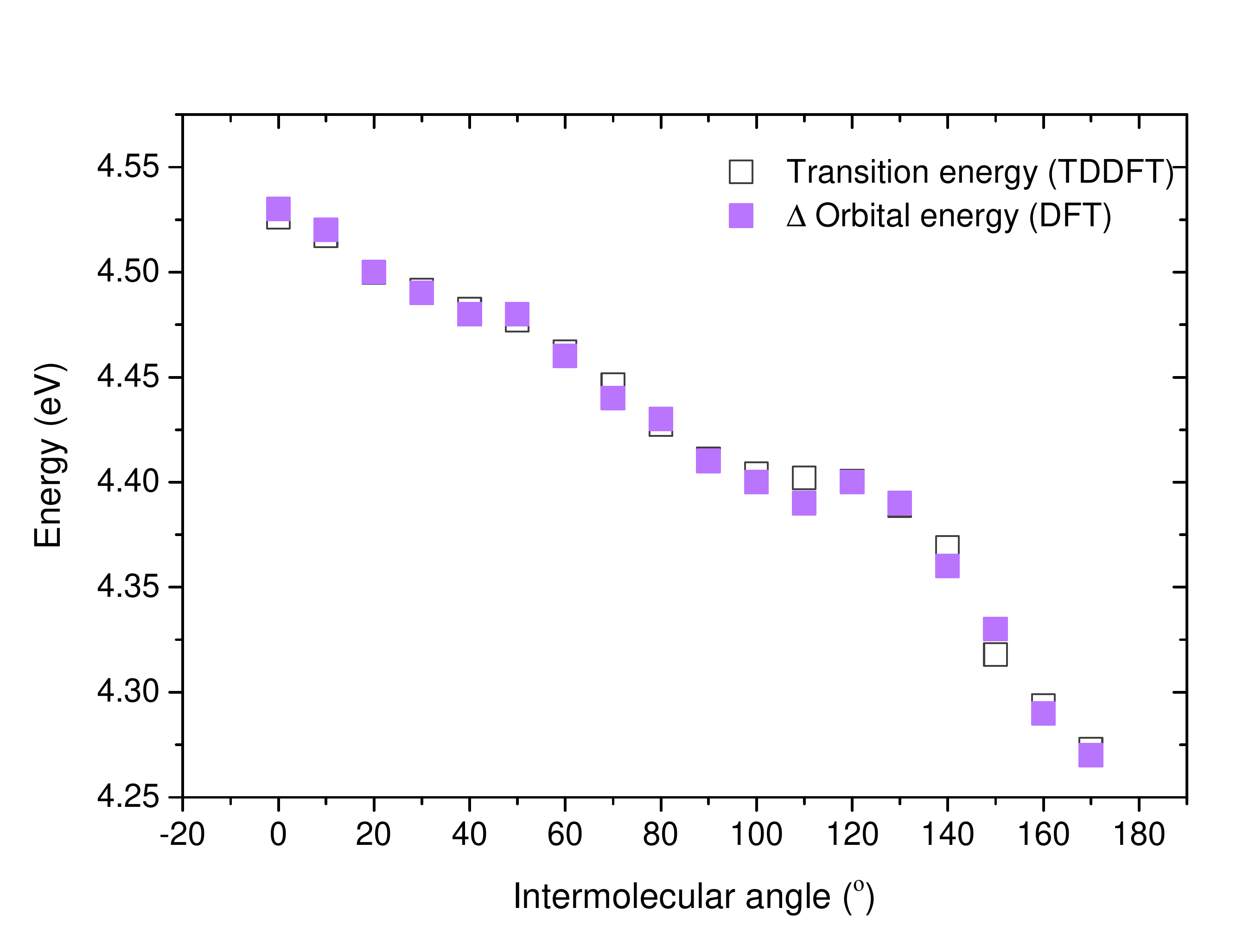}
        \caption{Excitation energy (TD-DFT) and Kohn Sham orbital energy difference at different intermolecular angles (first isosbestic region).}
        \label{fig:4ev}
    \end{figure}

    \item[(II)] Peak splitting region
    
        \begin{longtable}{ccccccc}
    \parbox{2cm}{Angle (${}^\circ$)} &
\parbox{2cm}{Excitation
Energy (eV) @TDDFT} &
\parbox{2cm}{Unoccupied
Orbital (UO) [eV] @DFT} &
\parbox{2cm}{Occupied Orbital (OO) [eV] @DFT} &
\parbox{2cm}{$\Delta$Orbital Energy
(UO-OO) [eV]}&
\parbox{2cm}{Mixing
Coefficient (\%)} &
\parbox{2cm}{Oscillator Strength}\\\hline
0 & 3.53554 & -4.48 & -8 & 3.52 & 88 & 0.02116\\
10 & 3.53959 & -4.49 & -8.01 & 3.52 & 80.3 & 0.02303\\
20 & 3.54321 & -4.5 & -8.02 & 3.52 & 79.9 & 0.0237\\
30 & 3.54441 & -4.5 & -8.03 & 3.53 & 79.5 & 0.02286\\
40 & 3.54074 & -4.51 & -8.03 & 3.52 & 79.2 & 0.01972\\
50 & 3.53279 & -4.52 & -8.03 & 3.51 & 75.1 & 0.01605\\
60 & 3.52431 & -4.52 & -8.03 & 3.51 & 48.7 & 0.01475\\
70 & 3.52124 & -4.52 & -8.03 & 3.51 & 53.2 & 0.017\\
80 & 3.52473 & -4.51 & -8.03 & 3.52 & 60.3 & 0.019\\
90 & 3.52747 & -4.5 & -8.03 & 3.53 & 60.1 & 0.01708\\
100 & 3.52658 & -4.5 & -8.03 & 3.53 & 58 & 0.01819\\
110 & 3.52239 & -4.51 & -8.03 & 3.52 & 47.3 & 0.01107\\
120 & 3.48626 & -4.52 & -8.02 & 3.5 & 60.3 & 0.01186\\
130 & 3.46751 & -4.53 & -8 & 3.47 & 68.6 & 7.98062E-4\\
140 & 3.43397 & -4.54 & -7.97 & 3.43 & 81 & 0.01315\\
150 & 3.40127 & -4.55 & -7.94 & 3.39 & 82.7 & 0.00588\\
160 & 3.37353 & -4.55 & -7.91 & 3.36 & 68.1 & 0.00158\\
170 & 3.32948 & -4.56 & -7.89 & 3.33 & 48.6 & 1.16965E-4\\\hline
    \caption{Angle dependence of the excitation energies and occupied and unoccupied orbitals for the peak-splitting at 3.6 eV.}

\end{longtable}

\newpage
    \begin{figure}
        \centering
        \includegraphics[width=0.7\textwidth]{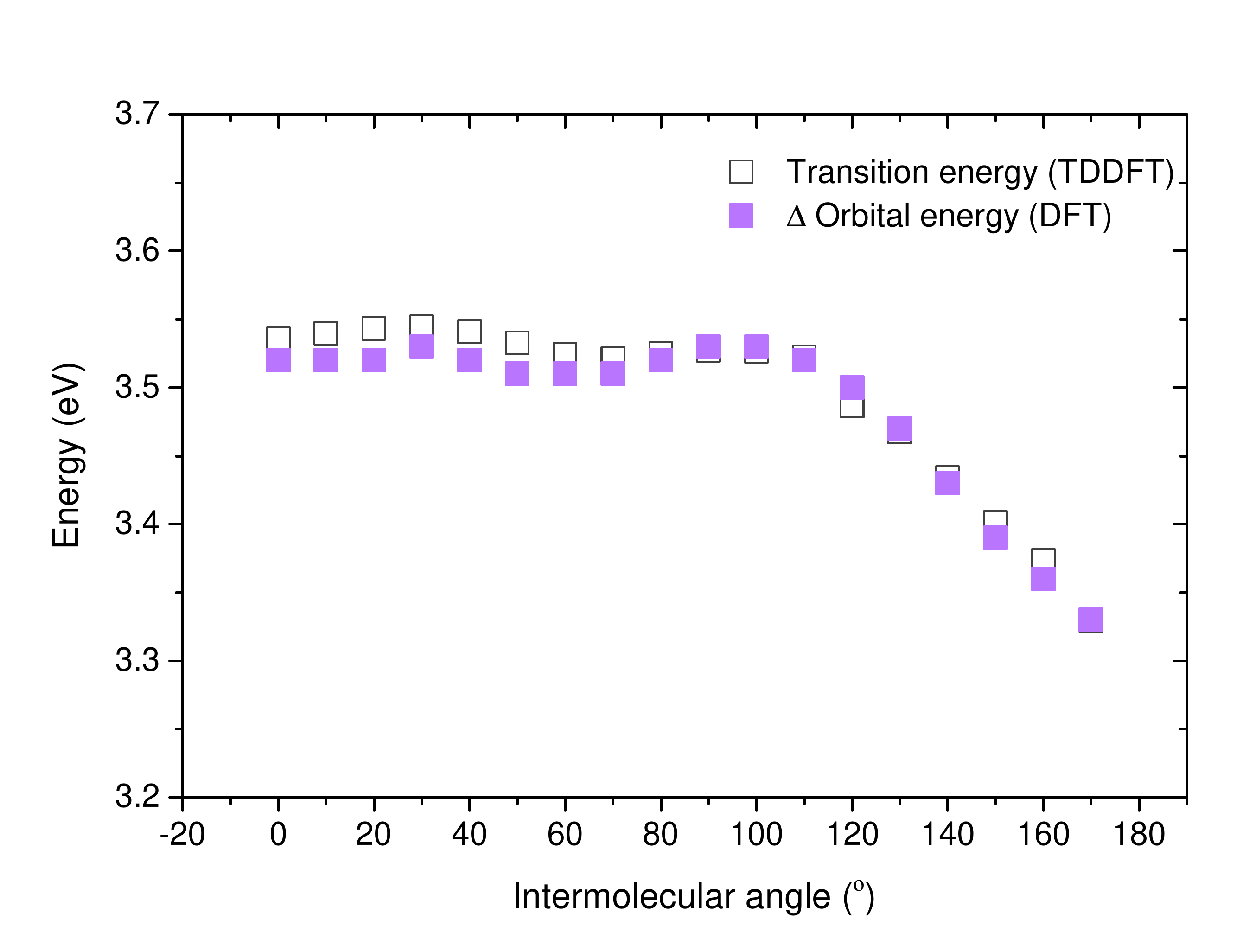}
        \caption{Excitation energy (TDDFT) and Kohn Sham orbital energy difference at different intermolecular angles (peak split region).}
        \label{fig:35ev}
    \end{figure}
    
    \item[(III)] Second Isosbectic point
    \begin{longtable}{ccccccc}
    \parbox{2cm}{Angle (${}^\circ$)} &
\parbox{2cm}{Excitation
Energy (eV) @TDDFT} &
\parbox{2cm}{Unoccupied
Orbital (UO) [eV] @DFT} &
\parbox{2cm}{Occupied Orbital (OO) [eV] @DFT} &
\parbox{2cm}{$\Delta$Orbital Energy
(UO-OO) [eV]}&
\parbox{2cm}{Mixing
Coefficient (\%)} &
\parbox{2cm}{Oscillator Strength}\\\hline
0 & 3.00897 & -3.36 & -6.36 & 3 & 57 & 1.21823E-4\\
10 & 3.01192 & -3.38 & -6.37 & 2.99 & 62 & 4.40075E-5\\
20 & 3.01608 & -3.38 & -6.38 & 3 & 65 & 9.08277E-6\\
30 & 3.02192 & -3.4 & -6.39 & 2.99 & 53.4 & 5.82045E-4\\
40 & 3.0249 & -3.39 & -6.41 & 3.02 & 51.2 & 6.39854E-5\\
50 & 3.03204 & -3.39 & -6.43 & 3.04 & 55.2 & 3.04819E-6\\
60 & 3.043 & -3.4 & -6.44 & 3.04 & 29.4 & 1.3827E-4\\
70 & 3.05324 & -3.4 & -6.44 & 3.04 & 41 & 2.1428E-5\\
80 & 3.04 & -3.4 & -6.44 & -3.04 & 33 & 2.01372E-4\\
90 & 3.03851 & -3.4 & -6.43 & 3.03 & 39.3 & 9.23179E-4\\
100 & 3.03386 & -3.41 & -6.42 & 3.01 & 37 & 0.00188\\
110 & 3.02784 & -3.42 & -6.41 & 2.99 & 25 & 0.00219\\
120 & 2.98179 & -3.42 & -6.4 & 2.98 & 58 & 3.68902E-5\\
130 & 2.97257 & -3.43 & -6.39 & 2.96 & 76 & 3.6898E-6\\
140 & 2.9603 & -3.45 & -6.39 & 2.94 & 79.2 & 7.58989E-5\\
150 & 2.94546 & -3.46 & -6.39 & 2.93 & 63 & 2.23839E-4\\
160 & 2.93369 & -3.47 & -6.38 & 2.91 & 70 & 2.78303E-4\\
170 & 2.92649 & -3.47 & -6.38 & 2.91 & 74.4 & 9.55178E-6\\\hline
\caption{Angle dependence of the excitation energies and occupied and unoccupied orbitals for the second isosbectic point.}
\end{longtable}
\newpage
\begin{figure}[h!]
    \centering
    \includegraphics[width=0.8\textwidth]{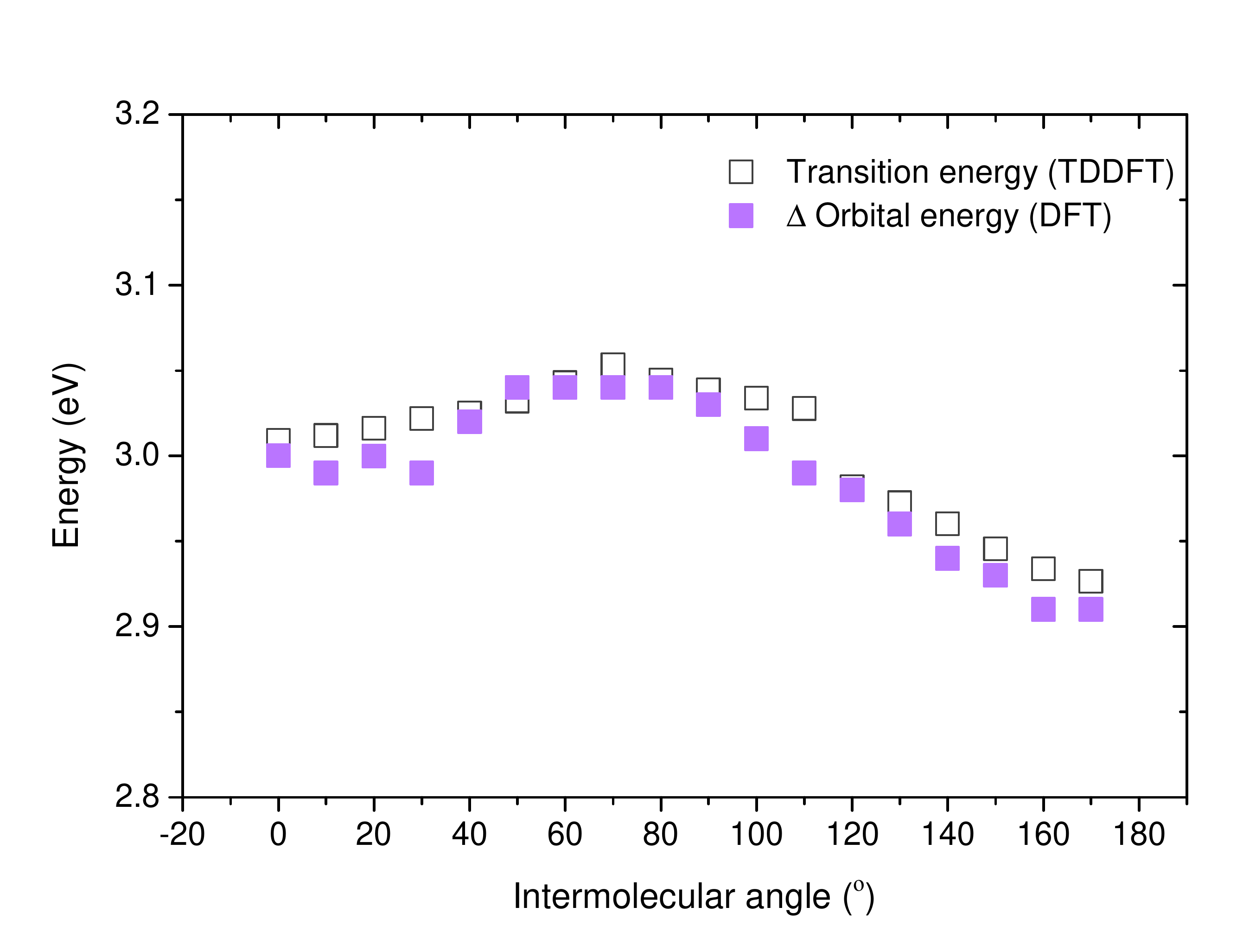}
    \caption{Excitation energy (TDDFT) and Kohn Sham orbital energy difference at different intermolecular angles (second isosbestic region).}
    \label{fig:3ev}
\end{figure}
\end{itemize}

\end{itemize}

\newpage

\subsection{Correlation between TD-DFT energies and orbital energies}
\begin{figure}[h!]
    \centering
    \includegraphics[width=0.8\textwidth]{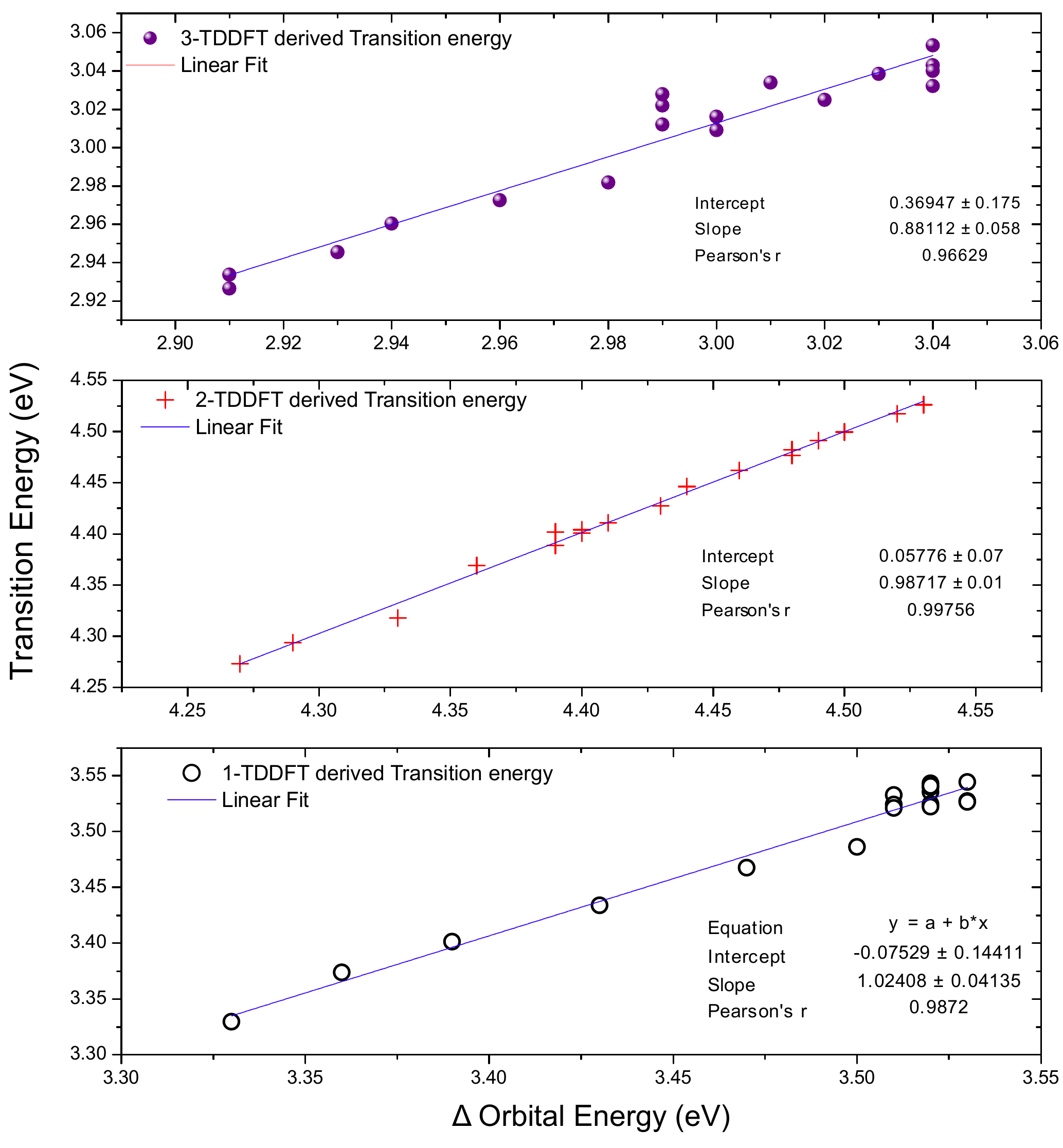}
    \caption{Correlation between transition energy (TDDFT) and Kohn Sham orbital energy difference for the three different transitions}
    \label{fig:corr}
\end{figure}

Figure~\ref{fig:corr} describes the correlation between the transition energies for the dimer structures at different intermolecular angles and the difference between the energies of the main contributing orbitals. The plot graphed by linear fitting shows a correlation (Pearsons’s R) atleast 97\% in each case.

\newpage
\subsection{Comparison with the Tracked Orbitals Estimated via DFT}

\begin{figure}
    \centering
    \includegraphics[width=0.7\columnwidth]{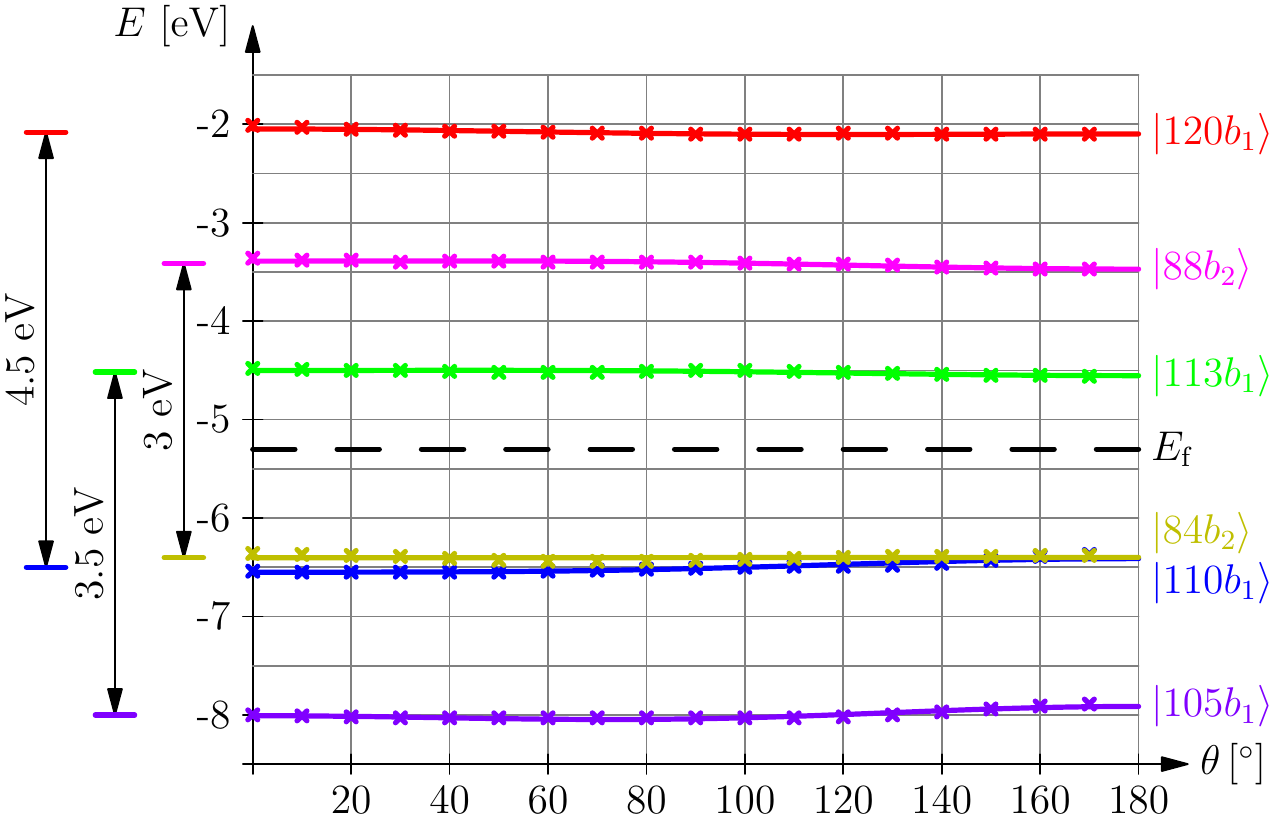}
    \caption{Variation of selected Kohn-Sham energies (see text) with dihedral angle. The DFT values are given by the dots in the highlighted $\theta$ range. Solid lines are fits according to Eq. (\ref{eq:fitKS}). Dashed line represents the position of the Fermi energy $E_{\rm f}$.
    }
    \label{fig:orbitals}
\end{figure}

We first study the impact of rotation on the energies of the  Kohn-Sham orbitals governing the TD-DFT-derived intense transitions as listed in Table 1 in SI.
The evolution of the orbital's energies are plotted in Fig.~\ref{fig:orbitals}.
The orbital energies have been fitted in good agreement to the orientational dependence via 
\begin{equation}
    \label{eq:fitKS}
    \varepsilon_i(\theta) = \varepsilon_{i} + \Delta \varepsilon_{i} f(e_i,\theta) \, ,
\end{equation}
with the fitting parameters $\Delta \varepsilon_{i}$ determining the strength of the interaction and the eccentricity of the considered Kohn-Sham orbital $e_i$, the energy of the monomer's orbital $\varepsilon_{i}$ and  relation~(16). The results are given in Table~\ref{tab_orb}.
\begin{table}[htb]
    \centering
    \begin{tabular}{c|ccc}
        orbital & eccentricity & fitting agreement & colour  \\\hline
 $\left|120b_1\right\rangle$ & 1.419 & 98\% & red \\
 $\left|88b_2\right\rangle$ & 0.610 & 91\% & magenta \\
 $\left|113b_1\right\rangle$ & 0.613 & 72\% & green \\
  $\left|84b_2\right\rangle$ & 0.969 & 79\% & yellow\\
 $\left|110b_1\right\rangle$ & 0.472 & 95\% & blue \\
$\left|105b_1\right\rangle$ & 0.818 & 90\% & purple 
    \end{tabular}
    \caption{Results of the fitting routine for the six tracked orbitals estimated via DFT simulations ($\left|\varphi_n\right\rangle$), the corresponding eccentricities of the orbitals, the fitting agreement, which is the coefficient of determination $R^2$, and colour for fig.~\ref{fig:orbitals}. }
    \label{tab_orb}
\end{table}

It can be observed that the single MOs fit quite well to the model~(\ref{eq:fitKS}). The deviation can be explain via the mismatch of the particle's inner-structure.

\end{document}